\documentclass[a4paper]{article}
\makeatletter
\def\@seccntformat#1{\@ifundefined{#1@cntformat}%
   {\csname the#1\endcsname\quad}  
   {\csname #1@cntformat\endcsname}
}
\let\oldappendix\appendix 
\renewcommand\appendix{%
    \oldappendix
    \newcommand{\section@cntformat}{\appendixname~\thesection\quad}
}
\makeatother

\usepackage{graphicx}

\usepackage{cite}
\usepackage[normalem]{ulem}

\usepackage[usenames,dvipsnames]{xcolor} 

\usepackage{amssymb,amsfonts,amsmath}
\usepackage[top=25truemm,bottom=25truemm,left=20truemm,right=20truemm]{geometry}
\usepackage{bm}
\usepackage{cases}
\usepackage[mathlines]{lineno}

\usepackage{setspace}

\title{Zero-determinant strategies in finitely repeated games}
\bigskip
\author{Genki Ichinose${}^{1}$ and Naoki Masuda${}^{2*}$
\ \\
\ \\
${}^{1}$
Department of Mathematical and Systems Engineering, Shizuoka University, \\3-5- 1 Johoku, Naka-ku, Hamamatsu, 432-8561, Japan\\
${}^{2}$
Department of Engineering Mathematics, University of Bristol, \\Merchant Venturers Building, Woodland Road, Clifton, Bristol BS8 1UB, United Kingdom\\
$^*$ Corresponding author (naoki.masuda@bristol.ac.uk)}

\begin{document}

\maketitle

\section*{Abstract}
Direct reciprocity is a mechanism for sustaining mutual cooperation in repeated social dilemma games, where a player would keep cooperation to avoid being retaliated by a co-player in the future. So-called zero-determinant (ZD) strategies enable a player to unilaterally set a linear relationship between the player's own payoff and the co-player's payoff regardless of the strategy of the co-player. In the present study, we analytically study zero-determinant strategies in finitely repeated (two-person) prisoner's dilemma games with a general payoff matrix. Our results are as follows. First, we present the forms of solutions that extend the known results for infinitely repeated games (with a discount factor $w$ of unity) to the case of finitely repeated games ($0<w<1$). Second, for the three most prominent ZD strategies, the equalizers, extortioners, and generous strategies, we derive the threshold value of $w$ above which the ZD strategies exist. Third, we show that the only strategies that enforce a linear relationship between the two players' payoffs are either the ZD strategies or unconditional strategies, where the latter independently cooperates with a fixed probability in each round of the game, proving a conjecture previously made for infinitely repeated games.

\section*{Keywords}
Prisoner's dilemma game; Cooperation; Direct reciprocity; Discount factor

\section{Introduction\label{sec:introduction}}

The prisoner's dilemma game models situations in which two individuals are involved in a social dilemma and each individual selects either cooperation (C) and defection (D) in the simplest setting. Although an individual obtains a larger payoff by selecting D regardless of the choice of the other individual, mutual defection, which is the unique Nash equilibrium of the game, yields a smaller benefit to both players than mutual cooperation does. We now know various mechanisms that enable mutual cooperation in the prisoner's dilemma game and other social dilemma games \cite{Nowak2006book,Sigmund2010book,Rand2013TrendsCognSci}, which inform us how cooperation is probably sustained in society of humans and animals and how to design cooperative organisations and society.

One of the mechanisms enabling mutual cooperation in social dilemma games is direct reciprocity, i.e., repeated interaction, in which the same two individuals play the game multiple times. An individual that defects would be retaliated by the co-player in the succeeding rounds. Therefore, the rational decision for both players in the repeated prisoner's dilemma game is to keep mutual cooperation if the number of iteration is sufficiently large \cite{Trivers1971QRevBiol,Axelrod1984book,Nowak2006book}. 
Generous tit-for-tat \cite{Nowak1992Nature-gtft} and win-stay lose-shift (often called Pavlov) \cite{Kraines1993TheorDecis,Nowak1993Nature} strategies are strong competitors in evolutionary dynamics of the repeated prisoner's dilemma game under noise, and a population composed of them realizes a high level of mutual cooperation.

In 2012, when the study of direct reciprocity seemed to be matured, Press and Dyson proposed a novel class of strategies in the repeated prisoner's dilemma game, called zero-determinant (ZD) strategies \cite{Press2012PNAS}. ZD strategies impose a linear relationship between the payoff obtained by a focal individual and its co-player regardless of the strategy that the co-player implements. A special case of the ZD strategies is the equalizer, with which the focal individual unilaterally determines the payoff that the co-player gains regardless of what the co-player does, within a permitted range of the co-player's payoff value (see \cite{Boerlijst1997AmMathMonth,Sigmund2010book} for the previous accounts for this strategy). As a different special case, the focal individual can set an ``extortionate'' share of the payoff that the individual gains as compared to the co-player's payoff. The advent of the ZD strategies has spurred new lines of investigations of direct reciprocity. They include the examination and extension of ZD strategies such as their evolution \cite{Stewart2012PNAS,Akin2012-2013arxiv,Adami2013NatComm,Hilbe2013PNAS,Hilbe2013PlosOne-zd,ChenZinger2014JTheorBiol,Szolnoki2014PhysRevE-zd,Szolnoki2014SciRep-zd,WuRong2014PhysRevE-zd,Hilbe2015JTheorBiol,LiuLi2015PhysicaA-zd,Xu2017PhysRevE-zd}, 
multiplayer games \cite{Hilbe2014PNAS-zd,Hilbe2015JTheorBiol,Pan2015SciRep-zd,Milinski2016NatComm,Stewart2016PNAS}, continuous action spaces \cite{Mcavoy2016PNAS,Milinski2016NatComm,Stewart2016PNAS,Mcavoy2017TheorPopulBiol}, alternating games \cite{Mcavoy2017TheorPopulBiol},
human reactions to computerized ZD strategies \cite{Hilbe2014NatComm,Wang2016NatComm-zd},
and human-human experiments \cite{Hilbe2016PlosOne,Milinski2016NatComm}.

Most of the aforementioned mathematical and computational studies of the ZD strategies have been conducted under the assumption of infinitely repeated games.
While mathematically more elegant and advantageous, finitely repeated games are more realistic than infinitely repeated games and comply with experimental studies.
In the present study, we examine the ZD strategies in the finitely repeated prisoner's dilemma game. There are a few studies that have investigated ZD strategies in finitely repeated games.
Hilbe and colleagues defined and mathematically characterized ZD strategies in finitely repeated games
\cite{Hilbe2015GamesEconBehav} (also see \cite{Hilbe2014NatComm}).
McAvoy and Hauert analyzed ZD strategies in the finitely repeated donation game (i.e., a special case of the prisoner's dilemma game) in a continuous strategy space \cite{Mcavoy2016PNAS,Mcavoy2017TheorPopulBiol}.
Given these studies, our main contributions in the present article are summarized as follows.
First, we derive expressions for ZD strategies in finitely repeated games that are straightforward extensions of those previously found for the infinitely repeated game. Second, for the three most studied ZD strategies, we derive the threshold discount factor (i.e., how likely the next round of the game occurs in the finitely repeated game) above which the ZD strategy can exist. Third, we prove that imposing a linear relationship between the two individuals' payoffs implies that the focal player takes either the ZD strategy defined for finitely repeated games \cite{Hilbe2015GamesEconBehav} or an unconditional strategy (e.g., unconditional cooperation and unconditional defection), proving the conjecture in \cite{Hilbe2013PlosOne-zd} in the case of finitely repeated games.

\section{Preliminaries}

In this section, we explain the finitely repeated prisoner's dilemma game, the strategies of interest (i.e., memory-one strategies), and the expected payoffs. More thorough discussion of them is found in Refs.~\cite{Nowak1995JMathBiol,Sigmund2010book,Hilbe2015GamesEconBehav}.

We consider the symmetric two-person prisoner's dilemma game whose payoff matrix is
given by
\begin{equation}
\bordermatrix{
 & {\rm C} & {\rm D} \cr
{\rm C} & R & S \cr
{\rm D} & T & P \cr}.
\label{eq:payoff}
\end{equation}
The entries represent the payoffs that the focal player, denoted by $X$, gains in a single round of a repeated game.  Each row and column represents the action of the focal player, $X$, and the co-player (denoted by $Y$), respectively. 
We assume that
\begin{equation}
T>R>P>S,
\label{eq:T>R>P>S}
\end{equation}
which dictates the prisoner's dilemma game. Both players obtain a larger payoff by selecting D than C because $T>R$ and $P>S$. We also assume that
\begin{equation}
2R>T+S,
\label{eq:2R>T+S}
\end{equation}
which guarantees that mutual cooperation is more beneficial than the two players alternating C and D in the opposite phase, i.e., CD, DC, CD, DC, $\ldots$, where the first and second letter represent the actions selected by $X$ and $Y$, respectively \cite{Rapoport1965book,Axelrod1984book}.
The two players repeat the game whose payoff matrix in each round is given by
Eq.~\eqref{eq:payoff}. A next round given the current round takes place with probability $w$ ($0< w < 1$), which is called the discount factor.

Consider two players $X$ and $Y$ that adopt memory-one strategies, with which they use only the outcome of the last round to decide the action to be submitted in the current round. A memory-one strategy is specified by a 5-tuple; $X$'s strategy is given by a combination of
\begin{equation}
\bm{p}=(p_{\rm CC}, p_{\rm CD}, p_{\rm DC}, p_{\rm DD})
\label{eq:def bm p}
\end{equation}
and $p_0$, where $0\le p_{\rm CC}, p_{\rm CD}, p_{\rm DC}, p_{\rm DD}, p_0\le 1$. In Eq.~\eqref{eq:def bm p}, $p_{\rm CC}$ is the conditional probability that $X$ cooperates when both $X$ and $Y$ cooperated in the last round, $p_{\rm CD}$ is the conditional probability that $X$ cooperates when $X$ cooperated and $Y$ defected in the last round, $p_{\rm DC}$ is the conditional probability that $X$ cooperates when $X$ defected and $Y$ cooperated in the last round, and $p_{\rm DD}$ is the conditional probability that $X$ cooperates when both $X$ and $Y$ defected in the last round. Finally, $p_0$ is the probability that $X$ cooperates in the first round. Similarly, $Y$'s strategy is specified by a combination of
\begin{equation}
\bm{q}=(q_{\rm CC}, q_{\rm CD}, q_{\rm DC}, q_{\rm DD})
\end{equation}
and the probability to cooperate in the first round, $q_0$, where $0\le q_{\rm CC}, q_{\rm CD}, q_{\rm DC}, q_{\rm DD}, q_0\le 1$.

We refer to the first round of the repeated game as round 0. Because both players have been assumed to use a memory-one strategy, the stochastic state of the two players in round $t$ ($t\ge 0$) is specified by
\begin{equation}
\bm v(t) = \left( v_{\rm CC}(t), v_{\rm CD}(t), v_{\rm DC}(t), v_{\rm DD}(t)\right),
\end{equation}
where $v_{\rm CC}(t)$ is the probability that both players cooperate in round $t$, $v_{\rm CD}(t)$ is the probability that $X$ cooperates and $Y$ defects in round $t$, and so forth. The normalization is given by
$v_{\rm CC}(t) + v_{\rm CD}(t) + v_{\rm DC}(t) + v_{\rm DD}(t) = 1$ ($t=0, 1, \ldots$).
The initial condition is given by
\begin{equation}
\bm v(0) = \left(p_0 q_0, p_0 (1-q_0), (1-p_0)q_0, (1-p_0)(1-q_0) \right).
\label{eq:x(0)}
\end{equation}
Because the expected payoff to player $X$ in round $t$ is given by
$\bm v(t) \bm{S}_X^{\top}$, where
\begin{equation}
\bm{S}_X = (R, S, T, P),
\end{equation}
the expected per-round payoff to player $X$ in the repeated game is given by
\begin{equation}
\pi_X = (1-w)\sum_{t=0}^\infty w^t \bm v(t) \bm{S}_X^{\top}.
\label{eq:pi_X preliminary}
\end{equation}
The transition-probability matrix for $\bm v(t)$ is given by
\begin{equation}
M=\begin{pmatrix}
p_{\rm CC}q_{\rm CC} & p_{\rm CC}(1-q_{\rm CC}) & (1-p_{\rm CC})q_{\rm CC} & (1-p_{\rm CC})(1-q_{\rm CC}) \\
p_{\rm CD}q_{\rm DC} & p_{\rm CD}(1-q_{\rm DC}) & (1-p_{\rm CD})q_{\rm DC} & (1-p_{\rm CD})(1-q_{\rm DC}) \\
p_{\rm DC}q_{\rm CD} & p_{\rm DC}(1-q_{\rm CD}) & (1-p_{\rm DC})q_{\rm CD} & (1-p_{\rm DC})(1-q_{\rm CD}) \\
p_{\rm DD}q_{\rm DD} & p_{\rm DD}(1-q_{\rm DD}) & (1-p_{\rm DD})q_{\rm DD} & (1-p_{\rm DD})(1-q_{\rm DD})
\end{pmatrix}.
\label{eq:M}
\end{equation}
By substituting
\begin{equation}
\bm v(t) = \bm v(0) M^t
\end{equation}
in Eq.~\eqref{eq:pi_X preliminary}, one obtains
\begin{align}
\pi_X =& (1-w) \bm v(0)\sum_{t=0}^\infty (wM)^t \bm{S}_X^{\top}\notag\\
=& (1-w)\bm v(0)(I-wM)^{-1} \bm{S}_X^{\top},
\label{eq:pi_X}
\end{align}
where $I$ is the $4\times 4$ identity matrix. Similarly, the expected per-round payoff to player $Y$ is given by
\begin{equation}
\pi_Y = (1-w)\bm v(0)(I-wM)^{-1} \bm{S}_Y^{\top},
\label{eq:pi_Y}
\end{equation}
where
\begin{equation}
\bm{S}_Y = (R, T, S, P).
\end{equation}

\section{Results}

We search player $X$'s strategies that impose a linear relationship between the two players' payoffs, i.e.,
\begin{equation}
\alpha \pi_X + \beta \pi_Y + \gamma = 0.
\label{eq:pi_X and pi_Y linear}
\end{equation}
When $\alpha\neq 0$, we set $\chi = -\beta/\alpha$ and $\kappa = -\gamma/(\alpha+\beta)$ to transform Eq.~\eqref{eq:pi_X and pi_Y linear} to
\begin{equation}
\pi_X - \kappa = \chi (\pi_Y - \kappa).
\label{eq:non-equalizer}
\end{equation}

\subsection{Equalizer}

\subsubsection{Expression}

By definition, the equalizer unilaterally sets the co-player's payoff, $\pi_Y$, to a constant value irrespectively of the co-player's strategy \cite{Boerlijst1997AmMathMonth,Sigmund2010book,Press2012PNAS}. To derive an expression for the equalizer strategies in the finitely repeated game, we proceed along the following idea:
If a strategy $\bm p$ ensures that the payoffs of the two players are on a horizontal line in the $\pi_X$-$\pi_Y$ space, irrespective of the co-player's strategy, then the payoffs must be on that horizontal line if the co-player uses unconditional cooperation or unconditional defection. Substituting the co-player's unconditional cooperation and unconditional defection into the payoff formulas gives necessary conditions imposed on $X$'s strategy. A straightforward computation then shows that these necessary conditions are in fact often sufficient; even if the co-player uses strategies that are not unconditional cooperation or defection, the two payoffs lie on the same line. We will use the same idea in section~\ref{sub:general cases} as well.

Because the equalizer is equivalent to $\alpha=0$ in Eq.~\eqref{eq:pi_X and pi_Y linear} and hence not covered by Eq.~\eqref{eq:non-equalizer}, we start by rewriting Eq.~\eqref{eq:pi_Y} as follows:
\begin{align}
\label{sy_a}
\pi_Y =& (1-w)\bm v(0)\bm u^{\rm eq}\notag\\
=& (1-w)\left( p_0 q_0, p_0(1-q_0), (1-p_0)q_0, (1-p_0)(1-q_0) \right)
\begin{pmatrix}
u^{\rm eq}_1\\ u^{\rm eq}_2\\ u^{\rm eq}_3\\ u^{\rm eq}_4
\end{pmatrix}\notag\\
=& (1-w) \left[ p_0 q_0 u^{\rm eq}_1+ p_0(1-q_0)u^{\rm eq}_2+(1-p_0)q_0 u^{\rm eq}_3+(1-p_0)(1-q_0)u^{\rm eq}_4 \right],
\end{align}
where
\begin{equation}
\label{eq:u^eq}
\bm{u}^{\rm eq} = \begin{pmatrix}
u^{\rm eq}_1\\ u^{\rm eq}_2\\ u^{\rm eq}_3\\ u^{\rm eq}_4
\end{pmatrix}
\equiv (I-wM)^{-1} \bm{S}_Y^{\top}.
\end{equation}
We denote $\bm u^{\rm eq}$ when $Y$'s strategy is $\bm{q}=(0, 0, 0, 0)$ by $\bm u^{\rm eq, 0000}$. Note that $\bm u^{\rm eq, 0000}$ is independent of the probability that $Y$ cooperates in the initial round, i.e., $q_0$.
We denote by $\pi_{Y, 0000}$ the payoff of $Y$ when $\bm{q}=(0, 0, 0, 0)$.
Similarly, we denote $\bm u^{\rm eq}$ when $Y$'s strategy is $\bm{q}=(1, 1, 1, 1)$ by $\bm u^{\rm eq, 1111}$
 and by $\pi_{Y, 1111}$ the payoff of $Y$ when $\bm{q}=(1, 1, 1, 1)$. The expressions of $\bm u^{\rm eq, 0000}$, $\pi_{Y, 0000}$, $\bm u^{\rm eq, 1111}$, and $\pi_{Y, 1111}$ are given in Appendix~\ref{sec:equalizer u and pi}.
If $X$ applies an equalizer strategy, $\pi_{Y, 0000}=\pi_{Y, 1111}$ must hold true regardless of $q_0$. Therefore,
we obtain
\begin{align}
& (1-w) \left[p_0 q_0 u^{\rm eq, 0000}_1+ p_0(1-q_0) u^{\rm eq, 0000}_2+(1-p_0)q_0 u^{\rm eq, 0000}_3+(1-p_0)(1-q_0) u^{\rm eq, 0000}_4 \right]\notag\\
=& (1-w)\left[p_0 q_0 u^{\rm eq, 1111}_1+ p_0(1-q_0) u^{\rm eq, 1111}_2+(1-p_0)q_0 u^{\rm eq, 1111}_3 + (1-p_0)(1-q_0) u^{\rm eq, 1111}_4\right],
\end{align}
which leads to
\begin{align}
& q_0 \left[ p_0(u^{\rm eq, 0000}_1- u^{\rm eq, 1111}_1)-p_0(u^{\rm eq, 0000}_2- u^{\rm eq, 1111}_2)+(1-p_0)(u^{\rm eq, 0000}_3 - u^{\rm eq, 1111}_3)-(1-p_0)(u^{\rm eq, 0000}_4- u^{\rm eq, 1111}_4)\right]\notag\\
+& \left[p_0(u^{\rm eq, 0000}_2- u^{\rm eq, 1111}_2)+(1-p_0)(u^{\rm eq, 0000}_4 - u^{\rm eq, 1111}_4)\right] = 0.
\label{eq:equalizer cnd when f_Y moves}
\end{align}
Equation~\eqref{eq:equalizer cnd when f_Y moves} must hold true for arbitrary $0\le q_0 \le 1$. Therefore, we obtain
\begin{align}
p_0(u_{1}^{\rm eq, 0000}- u_{1}^{\rm eq, 1111})+(1-p_0)(u_{3}^{\rm eq, 0000}- u_{3}^{\rm eq, 1111}) =& 0,
\label{eq:cnd equalizer A}\\
p_0(u_{2}^{\rm eq, 0000} - u_{2}^{\rm eq, 1111})+(1-p_0)(u_{4}^{\rm eq, 0000} - u_{4}^{\rm eq, 1111}) =& 0.
\label{eq:cnd equalizer B}
\end{align}
Combination of  Eqs.~\eqref{eq:u^eq}, \eqref{eq:cnd equalizer A} and \eqref{eq:cnd equalizer B} leads to the following necessary conditions:
\begin{align}
p_{\rm CD} =& \cfrac{p_{\rm CC} (T-P)-(\frac{1}{w}+p_{\rm DD} )(T-R)}{R-P},
\label{eq:final p_CD equalizer}\\
p_{\rm DC} =& \cfrac{(\frac{1}{w}-p_{\rm CC} )(P-S)+p_{\rm DD} (R-S)}{R-P},
\label{eq:final p_DC equalizer}
\end{align}
and $p_{\rm CC}$, $p_{\rm DD}$, and $p_0$ are arbitrary under the constraint $0\le p_{\rm CC}, p_{\rm CD}, p_{\rm DC}, p_{\rm DD}, p_0\le 1$. Equations~\eqref{eq:final p_CD equalizer} and \eqref{eq:final p_DC equalizer} extend the results previously obtained for $w=1$ \cite{Press2012PNAS}.

Surprisingly, Eqs.~\eqref{eq:final p_CD equalizer} and \eqref{eq:final p_DC equalizer} are also sufficient for $\bm p$ to be an equalizer strategy.
In other words, if a strategy of player $X$ satisfies Eqs.~\eqref{eq:final p_CD equalizer} and \eqref{eq:final p_DC equalizer}, then every co-player $Y$'s strategy, not restricted to unconditional cooperation or unconditional defection, yields the same payoff of $Y$. To verify this, we substitute
\begin{equation}
\bm{p}=\left(p_{\rm CC}, \cfrac{p_{\rm CC} (T-P)-(\frac{1}{w}+p_{\rm DD} )(T-R)}{R-P}, \cfrac{(\frac{1}{w}-p_{\rm CC} )(P-S)+p_{\rm DD} (R-S)}{R-P}, p_{\rm DD}\right)
\label{eq:final p equalizer}
\end{equation}
and $\bm{q}=(q_{\rm CC}, q_{\rm CD}, q_{\rm DC}, q_{\rm DD})$ in Eq.~\eqref{eq:u^eq} to obtain
\begin{equation}
\bm{u}^{\rm eq}= \frac{1}{(1-w)(1-wp_{\rm CC}+wp_{\rm DD})}
\begin{pmatrix}
w(1-p_{\rm CC})P+ (1-w+wp_{\rm DD})R\\
w(1-p_{\rm CC})P+ (1-w+wp_{\rm DD})R\\
(1-wp_{\rm CC})P + wp_{\rm DD} R\\
(1-wp_{\rm CC})P + wp_{\rm DD} R
\end{pmatrix},
\label{eq:bm a independent of q}
\end{equation}
which does not contain $\bm q$. By substituting Eq.~\eqref{eq:bm a independent of q} in
Eq.~\eqref{sy_a}, we obtain
\begin{equation}
\pi_Y = \cfrac{(1-p_0+wp_0- wp_{\rm CC})P + (p_0-wp_0+ wp_{\rm DD})R}
{1-wp_{\rm CC}+wp_{\rm DD}},
\label{eq:equalizer pi_Y value}
\end{equation}
which is independent of $\bm q$ and $q_0$.
Therefore, the set of the equalizer strategies is given by
Eq.~\eqref{eq:final p equalizer}, where $0\le p_{\rm CC}, p_{\rm CD}, p_{\rm DC}, p_{\rm DD}\le 1$,
combined with any $0\le p_0\le 1$.

It should be noted that an equalizer does not require any condition on $p_0$. However, Eq.~\eqref{eq:equalizer pi_Y value} indicates that the payoff that an equalizer enforces on the co-player, $\pi_Y$, depends on the value of $p_0$. Because Eq.~\eqref{eq:equalizer pi_Y value} is a weighted average of $P$ and $R$ with non-negative weights, an equalizer can impose any payoff value $\pi_Y$ such that $P\le \pi_Y\le R$.
If $P$ is enforced, it holds that $p_0-wp_0+wp_{\rm DD}=0$, and hence $p_{\rm DD} = p_0 = 0$. Therefore, the equalizer is a cautious strategy (i.e., never the first to cooperate) \cite{Hilbe2015GamesEconBehav}.
If $R$ is enforced, it holds that $1 - p_0 + wp_0 - wp_{\rm CC} = 0$, and hence $p_{\rm CC} = p_0 = 1$. Therefore, the equalizer is a nice strategy (i.e., never the first to detect) \cite{Hilbe2015GamesEconBehav}.
We remark that the equalizer is a ZD strategy for finitely repeated games as defined 
in Ref.~\cite{Hilbe2015GamesEconBehav} because it satisfies
Eq.~(31) of \cite{Hilbe2015GamesEconBehav} with $\alpha=0$.

\subsubsection{Minimum discount rate}

\begin{figure}[tb]
\begin{center}
\includegraphics[scale=1]{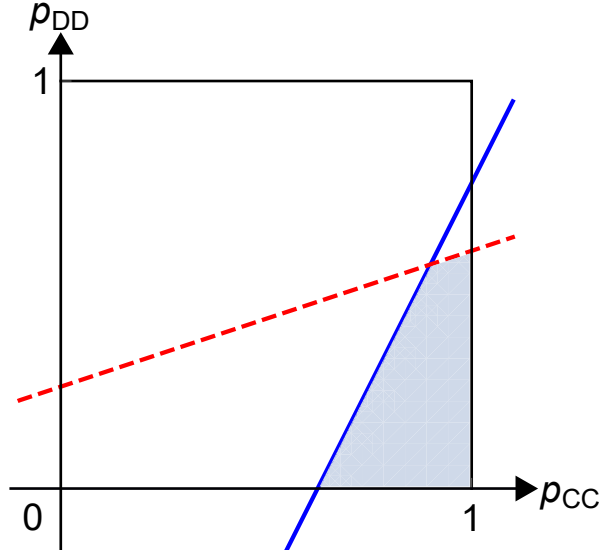}
\caption{Region in the $p_{\rm CC}$--$p_{\rm DD}$ space where the equalizer strategy exists (shaded region).
The border line of the half plane specified by Eqs.~\eqref{eq:cnd equalizer 3} and \eqref{eq:cnd equalizer 4} are shown by the solid and dashed lines, respectively. We set $R=3$, $T=5$, $S=0$, $P=1$, and $w=0.8$.}
\label{fig:schem equalizer cnd}
\end{center}
\end{figure}

In this section, we identify the condition for $w$ under which equalizer strategies exist.
Equation~\eqref{eq:final p equalizer} indicates that an equalizer strategy exists if and only if
\begin{align}
0 \le p_{\rm CC} (T-P)- \left(\frac{1}{w}+p_{\rm DD} \right)(T-R) \le R-P
\label{eq:cnd equalizer 1}
\end{align}
and
\begin{align}
0 \le \left(\frac{1}{w}-p_{\rm CC} \right)(P-S)+p_{\rm DD} (R-S) \le R-P
\label{eq:cnd equalizer 2}
\end{align}
for some $0\le p_{\rm CC}, p_{\rm DD}\le 1$. Note that we used Eq.~\eqref{eq:T>R>P>S}. Independently of $w$, any pair of $p_{\rm CC}$ and $p_{\rm DD}$ satisfies the second inequality of Eq.~\eqref{eq:cnd equalizer 1} and the first inequality of Eq.~\eqref{eq:cnd equalizer 2} because they are satisfied in the most stringent case, i.e., $p_{\rm CC}=1$ and $p_{\rm DD}=0$. The first inequality of Eq.~\eqref{eq:cnd equalizer 1} and the second inequality of Eq.~\eqref{eq:cnd equalizer 2} read
\begin{align}
p_{\rm DD} \le& \frac{T-P}{T-R} p_{\rm CC} - \frac{1}{w}
\label{eq:cnd equalizer 3}
\end{align}
and
\begin{align}
p_{\rm DD} \le& \frac{P-S}{R-S}p_{\rm CC} - \frac{1}{w}\frac{P-S}{R-S} + \frac{R-P}{R-S},
\label{eq:cnd equalizer 4}
\end{align}
respectively. Equations~\eqref{eq:cnd equalizer 3} and \eqref{eq:cnd equalizer 4} specify a $p_{\rm CC}$-$p_{\rm DD}$ region in the square $0\le p_{\rm CC}, p_{\rm DD}\le 1$, near the corner $(p_{\rm CC}, p_{\rm DD}) = (1, 0)$ (shaded region in Fig.~\ref{fig:schem equalizer cnd}). The feasible set $(p_{\rm CC}, p_{\rm DD})$ monotonically enlarges as $w$ increases. Therefore, we obtain the condition under which an equalizer exists by substituting $p_{\rm CC}=1$ and $p_{\rm DD}=0$ in Eqs.~\eqref{eq:cnd equalizer 3} and \eqref{eq:cnd equalizer 4}, i.e., 
\begin{equation}
w \ge w_{\rm c} \equiv \max \left(\frac{T-R}{T-P}, \frac{P-S}{R-S}\right).
\label{eq:cnd equalizer w}
\end{equation}
When $w = w_{\rm c}$, the unique equalizer strategy is given by $p_{\rm CC}=1$, $p_{\rm DD}=0$, and either $p_{\rm CD}$ or $p_{\rm DC}$ is equal to zero, depending on whether $(T-R)/(T-P)$ is larger than $(P-S)/(R-S)$ or vice versa.
The condition $w\ge (T-R)/(T-P)$ in Eq.~\eqref{eq:cnd equalizer w} coincides with that for the GRIM or tit-for-tat strategy to be stable against the unconditional defector \cite{Axelrod1984book}.

Equation~\eqref{eq:cnd equalizer w} is consistent with the result for the continuous donation game  \cite{Mcavoy2016PNAS}. Their result adapted to the case of two discrete levels of cooperation is $w_{\rm c} = c/b$, where $b$ and $c$ are the usual benefit and cost parameters in the donation game, respectively. We verify that Eq.~\eqref{eq:cnd equalizer w} with $R=b-c$, $T=b$,  $S=-c$, and $P=0$ yields $w_{\rm c}=c/b$.

\subsection{General cases\label{sub:general cases}}

All strategies but the equalizer in which a linear relationship is imposed between $\pi_X$ and $\pi_Y$ are given in the form of Eq.~\eqref{eq:non-equalizer}. In this section, we derive expressions of $X$'s strategy that realizes Eq.~\eqref{eq:non-equalizer}.

By substituting Eqs.~\eqref{eq:pi_X} and \eqref{eq:pi_Y} in Eq.~\eqref{eq:non-equalizer}, we obtain
\begin{equation}
\label{eq:zd2}
(1-w)\bm v(0)(I-wM)^{-1} \bm{S}_X^{\top} - \kappa = \chi \left[ (1-w)\bm v(0) (I-wM)^{-1} \bm{S}_Y^{\top} - \kappa \right].
\end{equation}
Equation~\eqref{eq:zd2} yields
\begin{equation}
\bm v(0)\left\{(1-w)(I-wM)^{-1} \left[\bm{S}_X^{\top} - \chi \bm{S}_Y^{\top} \right]+(\chi -1)\kappa \bm{1}\right\} = 0,
\label{eq:zd3}
\end{equation}
where
\begin{equation}
\bm{1}=\begin{pmatrix}
1\\ 1\\ 1\\ 1
\end{pmatrix}.
\end{equation}
We set
\begin{equation}
\label{eq:zd_u}
\bm u^{\rm zd}=
\begin{pmatrix}
u^{\rm zd}_1\\ u^{\rm zd}_2\\ u^{\rm zd}_3\\ u^{\rm zd}_4
\end{pmatrix}
\equiv (1-w)(I-wM)^{-1} \left[\bm{S}_X^{\top} - \chi \bm{S}_Y^{\top} \right]+(\chi -1)\kappa \bm{1}.
\end{equation}
Then, Eq.~\eqref{eq:zd3} is rewritten as
\begin{equation}
\label{eq:zd_xu}
\bm v(0) \bm u^{\rm zd} = 0,
\end{equation}
which is equivalent to
\begin{equation}
q_0 \left[ p_0 u^{\rm zd}_1-p_0 u^{\rm zd}_2+(1-p_0)u^{\rm zd}_3-(1-p_0)u^{\rm zd}_4 \right] + \left[ p_0 u^{\rm zd}_2 +(1-p_0)u^{\rm zd}_4 \right] = 0.
\label{eq:f' zd}
\end{equation}
Because Eq.~\eqref{eq:f' zd} must hold true irrespectively of $q_0$, we require
\begin{align}
  p_0 u^{\rm zd}_1+(1-p_0)u^{\rm zd}_3 =& 0,
  \label{eq:zd3-1}\\
  p_0 u^{\rm zd}_2+(1-p_0)u^{\rm zd}_4 =& 0.
  \label{eq:zd3-2}
\end{align}
Let us denote by $\bm u^{\rm zd, 0000}$ and $\bm u^{\rm zd, 1111}$ the vector $\bm u$ when $\bm{q}=(0, 0, 0, 0)$
and $\bm{q}=(1, 1, 1, 1)$, respectively. The expressions of $\bm u^{\rm zd, 0000}$ and $\bm u^{\rm zd, 1111}$ are given in Appendix~\ref{app:general cases}. By substituting $\bm u^{\rm zd, 0000}$ and $\bm u^{\rm zd, 1111}$ in
Eqs.~\eqref{eq:zd3-1} and \eqref{eq:zd3-2}, we obtain the four necessary conditions, Eqs.~\eqref{eq:zd cnd 1}, \eqref{eq:zd cnd 2}, \eqref{eq:zd cnd 3}, and \eqref{eq:zd cnd 4}, given in Appendix~\ref{app:general cases}.

If we assume $\kappa-S+\chi(T-\kappa) \neq 0$, we can rewrite Eq.~\eqref{eq:zd cnd 2} as
\begin{equation}
p_{\rm DD}=\cfrac{(1-w)p_0\left[(\chi-1)P +S - \chi T\right]  + (1-wp_{\rm CD})(\chi-1)(\kappa -P)}
{w \left[\kappa-S+\chi(T-\kappa)\right]}.
\label{eq:p_DD in terms of p_CD}
\end{equation}
If we assume $T-\kappa+\chi(\kappa-S) \neq 0$, we can rewrite Eq.~\eqref{eq:zd cnd 3} as
\begin{equation}
p_{\rm CC}=\cfrac{-(1-w)p_0\left[(\chi-1)R +T - \chi S\right]  + T - \chi S+(1+wp_{\rm DC})(\chi-1)\kappa - w p_{\rm DC}(\chi-1)R}
{w \left[T-\kappa+\chi(\kappa-S)\right]}.
\label{eq:p_CC in terms of p_DC}
\end{equation}
We will deal with the case $\kappa-S+\chi(T-\kappa) = 0$ or $T-\kappa+\chi(\kappa-S) = 0$ later in this section.

By substituting Eqs.~\eqref{eq:p_DD in terms of p_CD} and \eqref{eq:p_CC in terms of p_DC} in Eqs.~\eqref{eq:zd cnd 1}, we obtain an equation containing $p_{\rm CD}$, $p_{\rm DC}$, $p_0$, $\kappa$, and $\chi$ as unknowns. This equation can be factorized. By equating each of the two factors with 0,
we obtain two types of solutions. The one type of solution is given by
\begin{equation}
\bm{p}=\begin{pmatrix}
\cfrac{(1-w)p_0\left[(\chi-1)R +S - \chi T\right]  - (1-wp_{\rm CD})(\chi-1)R + \chi T -S - wp_{\rm CD}(\chi-1) \kappa}
{w \left[\kappa-S+\chi(T-\kappa)\right]}\\[1em]
p_{\rm CD}\\[1em]
\cfrac{-(1-w)p_0(\chi+1)(T-S) + (1-wp_{\rm CD})\left[(\chi-1)\kappa+T- \chi S\right]}
{w \left[\kappa-S+\chi(T-\kappa)\right]}\\[1em]
\cfrac{(1-w)p_0\left[(\chi-1)P +S - \chi T\right]  + (1-wp_{\rm CD})(\chi-1)(\kappa -P)}
{w \left[\kappa-S+\chi(T-\kappa)\right]}
\end{pmatrix}.
\label{eq:p zd 1}
\end{equation}
Equation~\eqref{eq:p zd 1} also satisfies Eq.~\eqref{eq:zd cnd 4}. To verify that Eq.~\eqref{eq:p zd 1} is sufficient, we substitute Eq.~\eqref{eq:p zd 1} in Eq.~\eqref{eq:zd_u} to obtain
\begin{equation}
\bm u^{\rm zd}=
\cfrac{(1 - w) \left[ S + (\chi-1) \kappa - \chi T\right]} {1 - wp_{\rm CD}-(1-w)p_0}
\begin{pmatrix}
1-p_0\\ 1-p_0\\ -p_0\\ -p_0
\end{pmatrix},
\label{eq:b zd sufficiency 1}
\end{equation}
which does not contain $\bm q$. Using Eqs.~\eqref{eq:x(0)} and \eqref{eq:b zd sufficiency 1}, we verify Eq.~\eqref{eq:zd_xu}.
Therefore, Eq.~\eqref{eq:p zd 1} is a set of strategies that impose the linear relationship between the payoff of the two players, i.e., Eq.~\eqref{eq:non-equalizer}.

The strategies given by Eq.~\eqref{eq:p zd 1} are ZD strategies for $w<1$ as defined in Ref.~\cite{Hilbe2015GamesEconBehav}, which is verified as follows. 
Assume that $\alpha\neq 0$ in Eq.~(31) of \cite{Hilbe2015GamesEconBehav} because $\alpha=0$ corresponds to the equalizer. Then, let us substitute $\alpha=\phi$, $\beta=-\phi \chi$, and $\gamma=\phi(\chi-1)\kappa$
in Eq.~(31) of \cite{Hilbe2015GamesEconBehav} without loss of generality. Note that 
this transformation is a bijection because (i) $\phi>0$ and (ii) either $\chi>1$ or $\chi<0$ is required (in the notation of Ref.~\cite{Hilbe2015GamesEconBehav}, $\phi>0$ and $\chi<1$ because their $\chi$ is defined as the reciprocal of our $\chi$).
Then, we obtain
\begin{equation}
w\bm{p}=\begin{pmatrix}
1-\phi (\chi-1)(R-\kappa)-(1-w)p_0\\[1em]
1+\phi\left[(\chi-1)\kappa-\chi T+S\right]-(1-w)p_0\\[1em]
\phi \left[(\chi-1)\kappa+T-\chi S\right]-(1-w)p_0\\[1em]
\phi(\chi-1)(\kappa-P)-(1-w)p_0
\end{pmatrix},
\label{eq:p zd 2}
\end{equation}
which is equivalent to Eq.~(33) of \cite{Hilbe2015GamesEconBehav}.
Equation~\eqref{eq:p zd 2} combined with
\begin{equation}
\label{phi_zd}
\phi = \cfrac{1-wp_{\rm CD}-p_0 (1-w)}{\kappa-S+\chi(T-\kappa)} 
\end{equation}
is equivalent to Eq.~\eqref{eq:p zd 1}. It should also be noted that Eq.~\eqref{eq:p zd 2} extends 
Eq.~(9) of \cite{ChenZinger2014JTheorBiol}, which has been obtained for $w=1$, to general $w$, $R$, and $P$ values.

The other type of solution that we obtain by substituting Eqs.~\eqref{eq:p_DD in terms of p_CD} and \eqref{eq:p_CC in terms of p_DC} in Eq.~\eqref{eq:zd cnd 1} is given by
\begin{equation}
p_0 [T-R+\chi(R-S)]=T-\kappa+\chi(\kappa-S).
\label{eq:p_0 cnd 1}
\end{equation}
Substitution of Eqs.~\eqref{eq:p_DD in terms of p_CD} and \eqref{eq:p_CC in terms of p_DC} in Eq.~\eqref{eq:zd cnd 4} yields either Eq.~\eqref{eq:p zd 1} or
\begin{equation}
p_0 [P-S+\chi(T-P)]=(\chi-1)(\kappa-P).
\label{eq:p_0 cnd 2}
\end{equation}
The combination of Eqs.~\eqref{eq:p_0 cnd 1} and \eqref{eq:p_0 cnd 2} is equivalent to that of
\begin{equation}
\kappa = p_0^2 R + p_0 (1 - p_0) (T + S) + (1 - p_0)^2 P
\label{eq:kappa rrrr}
\end{equation}
and
\begin{equation}
\chi = -\cfrac{(1 - p_0) (T - P) + p_0 (R - S)}{(1 - p_0) (P - S) + p_0 (T - R)}.
\label{eq:chi rrrr}
\end{equation}
However, Eqs.~\eqref{eq:p_DD in terms of p_CD}, \eqref{eq:p_CC in terms of p_DC},
\eqref{eq:kappa rrrr}, and \eqref{eq:chi rrrr} do not provide a sufficient condition for Eq.~\eqref{eq:zd_xu} to hold true for arbitrary $\bm q$ and $q_0$. Therefore, we additionally consider
the vector $\bm u^{\rm zd}$ when $\bm{q}=(1, 0, 0, 0)$
and $\bm{q}=(0, 0, 0, 1)$, which we denote by $\bm u^{\rm zd, 1000}$ and $\bm u^{\rm zd, 0001}$, respectively. 
The calculations shown in Appendix~\ref{sec:rrrrr calc} lead to
\begin{equation}
p_0 = p_{\rm CC} = p_{\rm CD} = p_{\rm DC} = p_{\rm DD}\quad (0\le p_0\le 1).
\label{eq:rrrrr}
\end{equation}

To verify that the unconditional strategies given by Eq.~\eqref{eq:rrrrr} are a sufficient condition for Eq.~\eqref{eq:non-equalizer} to hold true for arbitrary $\bm q$ and $q_0$, we substitute Eqs.~\eqref{eq:kappa rrrr}, \eqref{eq:chi rrrr}, and \eqref{eq:rrrrr} in Eq.~\eqref{eq:zd_u} to obtain
\begin{equation}
\bm u^{\rm zd}=
\cfrac{(1 - w) (T - S)} {-(1 - p_0) P + S + p_0 (R - S - T)}
\begin{pmatrix}
-(1 - p_0) [-(1 - p_0) P + (1 + p_0) R - p_0 (T + S)]\\ -(1 - p_0) [-(2 - p_0) P + T + S + p_0 (R - S - T)]\\ p_0 [-(1 - p_0) P + (1 + p_0) R - p_0 (T + S)]\\ p_0 [-(2 - p_0) P + T + S + p_0 (R - S - T)]
\end{pmatrix},
\label{eq:sufficiency 2}
\end{equation}
which does not contain $\bm q$. Using Eqs.~\eqref{eq:x(0)} and \eqref{eq:sufficiency 2}, we verify Eq.~\eqref{eq:zd_xu}. The unconditional strategy given by Eq.~\eqref{eq:rrrrr} is not a ZD strategy in the sense of \cite{Hilbe2015GamesEconBehav} unless $R+P=T+S$ (Appendix~\ref{sec:rrrrr is not ZD}), which is the same condition as that for the infinitely repeated game \cite{Hilbe2013PlosOne-zd}.

The obtained solution, i.e., Eq.~\eqref{eq:rrrrr} combined with Eqs.~\eqref{eq:kappa rrrr} and \eqref{eq:chi rrrr}, is equivalent to the previously derived solution for $w=1$ \cite{Hilbe2013PlosOne-zd}. This set of solutions contains the unconditional cooperator and unconditional defector as special cases, and always realizes $\chi<0$ (Eq.~\eqref{eq:chi rrrr}).

When $\kappa-S+\chi(T-\kappa) = 0$ or $T-\kappa+\chi(\kappa-S) = 0$,
the calculations shown in Appendices~\ref{sec:denom=0 case 1} and \ref{sec:denom=0 case 2} reveal the following three types of solutions: (i) a subset of the ZD strategies given by Eq.~\eqref{eq:p zd 1} (Appendix~\ref{sec:(D)}), (ii) a subset of the strategies given by Eq.~\eqref{eq:rrrrr} (Appendices~\ref{sec:(A)}, \ref{sec:(B)}, and \ref{sec:(D)}), and (iii) the set of strategies given by
\begin{equation}
\bm p = \left(p_{\rm CC}, \; 1, \; \cfrac{w p_{\rm CC} (\chi + 1) (\kappa - T) - w [R - (\chi + 1) T + \chi \kappa] - (\kappa - R)}{w(\kappa-R)}, \; \cfrac{w p_{\rm CC} (\kappa - P) - w (R - P) - (\kappa - R)}{w(\kappa-R)} \right), p_0=1,
\label{eq:sol-B-1} 
\end{equation}
where $0\le p_{\rm CC}\le 1$ and $\kappa \neq R$ (Appendix~\ref{sec:(B)}). 
Although Eq.~\eqref{eq:sol-B-1} is a sufficient condition and the resulting solutions are distinct from those given by Eq.~\eqref{eq:p zd 1}, in fact Eq.~\eqref{eq:sol-B-1} yields $\chi<0$ (Appendix~\ref{sec:(B)}).

To summarize, the set of $X$'s strategies that enforce Eq.~\eqref{eq:non-equalizer} is the union of the strategies given by the ZD strategies, Eq.~\eqref{eq:p zd 1}, and the non-ZD unconditional strategies, Eq.~\eqref{eq:rrrrr}.
In the next sections, we examine two special cases, which have been studied in the literature, and derive $w_{\rm c}$ in each case.

\subsection{Extortioner}

\subsubsection{Expression}

The extortioner is defined as a strategy that enforces an extortionate share of payoffs larger than $P$ 
\cite{Press2012PNAS}. We obtain the extortioner by setting $\kappa=P$ in Eq.~\eqref{eq:non-equalizer}. By setting $\kappa = P$ in Eq.~\eqref{eq:p zd 1}, we obtain
\begin{equation}
\bm{p}=\begin{pmatrix}
\cfrac{(1-w)p_0\left[(\chi-1)R +S - \chi T\right]  - (1-wp_{\rm CD})(\chi-1)R + \chi T -S - wp_{\rm CD}(\chi-1) P}
{w \left[P-S+\chi(T-P)\right]}\\[1em]
p_{\rm CD}\\[1em]
\cfrac{-(1-w)p_0(\chi+1)(T-S) + (1-wp_{\rm CD})\left[(\chi-1)P+T- \chi S\right]}
{w \left[P-S+\chi(T-P)\right]}\\[1em]
- \cfrac{(1-w)p_0}{w}
\end{pmatrix}.
\label{eq:p extortioner 1}
\end{equation}
Because $p_{\rm DD} = -(1-w)p_0/w \geq 0$ and $w<1$, we obtain $p_0=0$ and $p_{\rm DD}=0$, which is consistent with the previously obtained result \cite{Hilbe2015GamesEconBehav}. Therefore, the extortioner is never the first to cooperate and hence a so-called cautious strategy \cite{Hilbe2015GamesEconBehav}. By setting $p_0=0$ in Eq.~\eqref{eq:p extortioner 1}, we obtain
\begin{equation}
\label{ext6}
\bm{p}=\begin{pmatrix}\cfrac{ - w p_{\rm CD} (\chi-1)P -(1-wp_{\rm CD})(\chi-1)R - S + \chi T}
{w \left[P-S+\chi(T-P)\right]}\\[1em]
p_{\rm CD}\\[1em]
\cfrac{(1-wp_{\rm CD})\left[(\chi-1)P+T-\chi S\right]}
{w \left[P-S+\chi(T-P)\right]}\\[1em]
0
\end{pmatrix}.
\end{equation}

\subsubsection{Minimum discount rate}

By setting $\kappa=P$ and $p_0=0$ in Eq.~\eqref{eq:p zd 2}, we obtain
\begin{equation}
w\bm{p}=\begin{pmatrix}
1-\phi (\chi-1)(R-P)\\[1em]
1+\phi\left[(\chi-1)P-\chi T+S\right]\\[1em]
\phi \left[(\chi-1)P+T-\chi S\right]\\[1em]
0
\end{pmatrix}.
\label{eq:extortioner bm p with phi and chi}
\end{equation}
Because $p_{\rm CC}\le 1$ and $w<1$, Eq.~\eqref{eq:extortioner bm p with phi and chi} implies that $\phi(\chi-1)>0$ must hold true. We consider
the case $\phi>0$ and $\chi>1$ in this section. We can exclude the case $\phi<0$ and $\chi<1$ because a strategy with $\chi<0$ is not considered as an extortionate strategy  \cite{Press2012PNAS, Hilbe2013PNAS, Hilbe2013PlosOne-zd, ChenZinger2014JTheorBiol, Xu2017PhysRevE-zd, Pan2015SciRep-zd, Mcavoy2016PNAS, Mcavoy2017TheorPopulBiol, Hilbe2015GamesEconBehav, Stewart2013PNAS} and $\chi<1$ implies $\chi<0$ (Appendix~\ref{app:extortioner chi<0}).

When $\phi>0$, the application of $0\le p_{\rm CC}, p_{\rm CD}, p_{\rm DC}\le 1$ to Eq.~\eqref{eq:extortioner bm p with phi and chi} yields
\begin{align}
\frac{(\chi-1)\frac{R-P}{P-S}} {\frac{1}{w}} \le& \frac{1}{\phi} \le \frac{(\chi-1)\frac{R-P}{P-S}} {\frac{1}{w}-1},
\label{eq:1/phi 1}\\
\frac{1+\chi\frac{T-P}{P-S}} {\frac{1}{w}} \le& \frac{1}{\phi} \le \frac{1+\chi\frac{T-P}{P-S}} {\frac{1}{w}-1},
\label{eq:1/phi 2}\\
\chi + \frac{T-P}{P-S} \le& \frac{1}{\phi}.
\label{eq:1/phi 3}
\end{align}
The condition under which a positive $\phi$ value that satisfies Eqs.~\eqref{eq:1/phi 1}, \eqref{eq:1/phi 2}, and \eqref{eq:1/phi 3} exists is given by
\begin{align}
\frac{(\chi-1)\frac{R-P}{P-S}} {\frac{1}{w}} \le& \frac{1+\chi\frac{T-P}{P-S}} {\frac{1}{w}-1},
\label{eq:phi exists 1}\\
\frac{1+\chi\frac{T-P}{P-S}} {\frac{1}{w}} \le& \frac{(\chi-1)\frac{R-P}{P-S}} {\frac{1}{w}-1},
\label{eq:phi exists 2}\\
\chi + \frac{T-P}{P-S} \le& \frac{(\chi-1)\frac{R-P}{P-S}} {\frac{1}{w}-1},
\label{eq:phi exists 3}\\
\chi + \frac{T-P}{P-S} \le& \frac{1+\chi\frac{T-P}{P-S}} {\frac{1}{w}-1}.
\label{eq:phi exists 4}
\end{align}
Equation~\eqref{eq:phi exists 1} is always satisfied. Equations~\eqref{eq:phi exists 2}, \eqref{eq:phi exists 3}, and \eqref{eq:phi exists 4} yield
\begin{align}
\chi\left[w(T-P) - (T-R)\right] \ge& R-S - w(P-S),
\label{eq:w cnd extortion 1}\\
%
\chi\left[w(T-S)-(P-S)\right] \ge& T-P - w(T-S),
\label{eq:w cnd extortion 2}\\
%
\chi\left[w(R-S)-(P-S)\right] \ge& T-P - w(T-R),
\label{eq:w cnd extortion 3}
\end{align}
respectively.

The left-hand side of Eq.~\eqref{eq:w cnd extortion 2} is always larger than that of Eq.~\eqref{eq:w cnd extortion 3}, and the right-hand side of Eq.~\eqref{eq:w cnd extortion 2} is always smaller than that of Eq.~\eqref{eq:w cnd extortion 3}. Therefore, Eq.~\eqref{eq:w cnd extortion 2} is satisfied if Eq.~\eqref{eq:w cnd extortion 3} is satisfied.
The right-hand sides of Eqs.~\eqref{eq:w cnd extortion 1} and \eqref{eq:w cnd extortion 3} are positive. Therefore, $w(T-P)-(T-R)>0$ and $w(T-S)-(P-S)>0$ are required for $\chi$ to be positive. 
On the other hand, if $w(T-P)-(T-R)>0$ and $w(T-S)-(P-S)>0$, Eqs.~\eqref{eq:w cnd extortion 1} and \eqref{eq:w cnd extortion 3} guarantee that $\chi>1$ and that a $\chi (>1)$ value exists.
Therefore, an extortioner with $\chi>1$ exists if and only if $w > w_{\rm c}$, where
the $w_{\rm c}$ value coincides with that for the equalizer; it is given by Eq.~\eqref{eq:cnd equalizer w}.
Under $w > w_{\rm c}$,  Eqs.~\eqref{eq:w cnd extortion 1} and \eqref{eq:w cnd extortion 3} imply
\begin{equation}
\chi \ge \chi_{\rm c}(w) \equiv \max \left( 
\frac{R-S - w(P-S)}{w(T-P) - (T-R)}, \frac{T-P - w(T-R)}{w(R-S)-(P-S)} \right).
\label{eq:extortioner chi range}
\end{equation}
Equation~\eqref{eq:extortioner chi range} gives the range of $\chi$ values for which the extortioner strategy exists. The conditions for the existence of an extortionate strategy are easier to satisfy for large $w$ in the sense that $\chi_{\rm c}(w)$ monotonically decreases as $w$ increases. In particular, we obtain $\lim_{w\to w_{\rm c}+0} \chi_{\rm c}(w) = \infty$ and $\lim_{w\to 1} \chi_{\rm c}(w) = 1$.

For a given $\chi$ value, the substitution of $R=b-c$, $T=b$, $S=-c$, and $P=0$ in Eqs.~\eqref{eq:cnd equalizer w} yields
\begin{equation}
w_{\rm c} = \frac{\chi c + b}{\chi b + c},
\end{equation}
which is consistent with Eq.~(7) of \cite{Mcavoy2016PNAS}.


\subsection{Generous strategy}

\subsubsection{Expression}

The generous strategy, also called compliers, is defined as a strategy that yields a larger shortfall from the mutual cooperation payoff $R$ for the player as compared to that for the co-player \cite{Stewart2012PNAS,Stewart2013PNAS,Hilbe2013PlosOne-zd}.
We obtain the generous strategy by setting $\kappa=R$ in Eq.~\eqref{eq:non-equalizer}. By setting $\kappa = R$ in Eq.~\eqref{eq:p zd 1}, we obtain
\begin{equation}
\bm{p}=\begin{pmatrix}\cfrac{1-p_0(1-w)}{w} \\[1em]
p_{\rm CD} \\[1em]
\cfrac{-(1-w)p_0(\chi+1)(T-S) + (1-wp_{\rm CD})\left[(\chi-1)R+T- \chi S\right]}
{w \left[R-S+\chi(T-R)\right]}\\[1em]
\cfrac{(1-w)p_0\left[(\chi-1)P +S - \chi T\right]  + (1-wp_{\rm CD})(\chi-1)(R -P)}
{w \left[R-S+\chi(T-R)\right]}
\end{pmatrix}.
\label{eq:p generous 1}
\end{equation}
Because $p_{\rm CC} = \left[1-(1-w)p_0 \right]/w \leq 1$, we obtain $p_0=1$ and $p_{\rm CC}=1$,
which is consistent with the previously obtained result \cite{Hilbe2015GamesEconBehav}.
Therefore, the generous strategy is never the first to detect and hence a so-called nice strategy \cite{Axelrod1984book,Hilbe2015GamesEconBehav}.
By setting $p_0=1$ in Eq.~\eqref{eq:p generous 1}, we obtain
\begin{align}
\bm{p} =& \begin{pmatrix}
1\\[10pt]
p_{\rm CD} \\[10pt]
\cfrac{-(1-w)(\chi+1)(T-S) + (1-wp_{\rm CD})\left[(\chi-1)R+T- \chi S\right]}
{w \left[R-S+\chi(T-R)\right]}\\ \\
\cfrac{(1-w)\left[(\chi-1)P +S - \chi T\right]  + (1-wp_{\rm CD})(\chi-1)(R -P)}
{w \left[R-S+\chi(T-R)\right]}
\end{pmatrix}\notag\\
=&
\begin{pmatrix}
1\\[10pt]
p_{\rm CD}\\[10pt]
1-\frac{1}{w} + \frac{(1-p_{\rm CD}) \left[(\chi-1) R + T - \chi S\right]}
{R-S + \chi(T-R)}\\[10pt]
1-\frac{1}{w} + \frac{(1 - p_{\rm CD}) (\chi-1) (R-P)}
{R-S + \chi(T-R)}
\end{pmatrix}.
\label{eq:generous soln}
\end{align}

\subsubsection{Minimum discount rate}

\begin{figure}[tb]
\begin{center}
\includegraphics[scale=1]{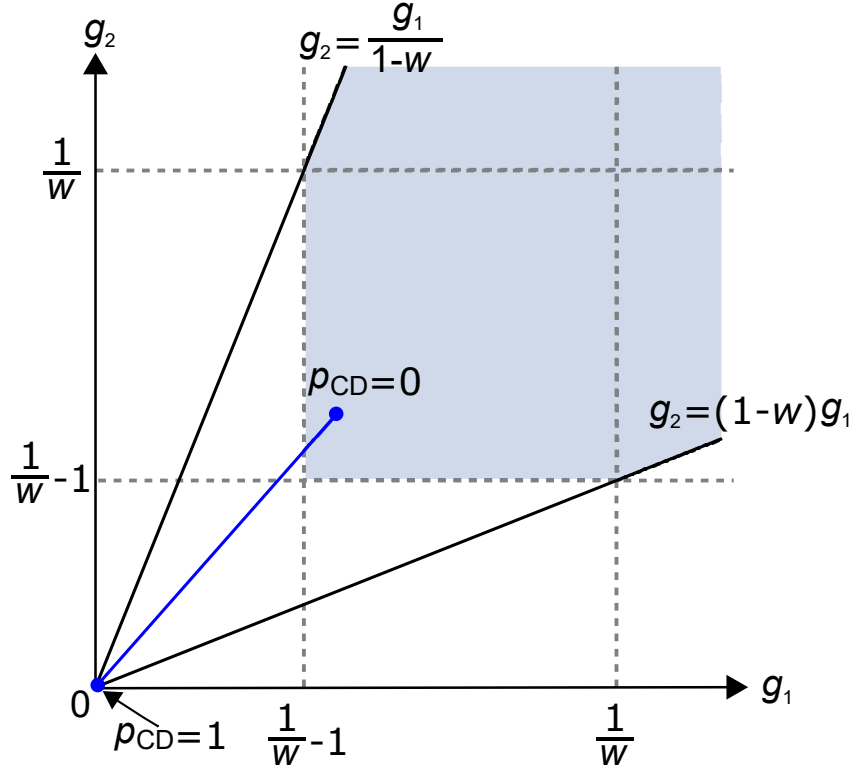}
\caption{Region in the $g_1$--$g_2$ space where the generous strategy exists (shaded region). If $(g_1, g_2)$ is located in this region (e.g., filled circle labeled $p_{\rm CD}=0$), the square given by $1/w-1\le g_1, g_2\le 1/w$ intersects the line segment connecting the assumed ($g_1, g_2$) and the origin. Note that any point on the line segment is realized by the solution by a value of $p_{\rm CD}$ (Eqs.~\eqref{eq:generous cnd g_1 with p_CD} and \eqref{eq:generous cnd g_2 with p_CD}).}
\label{fig:schem generous cnd}
\end{center}
\end{figure}

By applying $0\le p_{\rm DC}, p_{\rm DD}\le 1$ to Eq.~\eqref{eq:generous soln}, we obtain
\begin{align}
\frac{1}{w}-1 \le (1-p_{\rm CD})g_1 \le \frac{1}{w},
\label{eq:generous cnd g_1 with p_CD}\\
\frac{1}{w}-1\le (1-p_{\rm CD})g_2 \le \frac{1}{w},
\label{eq:generous cnd g_2 with p_CD}
\end{align}
where
\begin{align}
g_1 \equiv& \frac{(\chi-1)R+T-\chi S} {R-S+\chi(T-R)},\\
g_2 \equiv& \frac{(\chi-1)(R-P)} {R-S+\chi(T-R)}.
\end{align}
The necessary and sufficient condition for $0\le p_{\rm CD}\le 1$ that satisfies Eqs.~\eqref{eq:generous cnd g_1 with p_CD} and \eqref{eq:generous cnd g_2 with p_CD} to exist is given by (Fig.~\ref{fig:schem generous cnd})
\begin{align}
g_1 \ge& \frac{1}{w}-1,
\label{eq:generous cnd g_1}\\
g_2 \ge& \frac{1}{w}-1,
\label{eq:generous cnd g_2}\\
1-w \le& \frac{g_2}{g_1} \le \frac{1}{1-w}.
\label{eq:generous cnd g_2/g_1}
\end{align}

In the remainder of this section, we assume $\chi\ge0$, which a generous strategy requires \cite{Hilbe2013PlosOne-zd,ChenZinger2014JTheorBiol, Mcavoy2016PNAS, Mcavoy2017TheorPopulBiol, Hilbe2015GamesEconBehav, Stewart2013PNAS}, and examine the conditions given by Eqs.~\eqref{eq:generous cnd g_1}, \eqref{eq:generous cnd g_2}, and \eqref{eq:generous cnd g_2/g_1}. For mathematical interests, the analysis of the minimum discount rate for $\chi < 0$ is presented in Appendix~\ref{app:generous chi<0}.
First, because ${\rm d}g_1/{\rm d}\chi > 0$, which one can derive using Eq.~\eqref{eq:2R>T+S}, and $g_1$ is continuous for $\chi\ge 0$,
Eq.~\eqref{eq:generous cnd g_1} is equivalent to
\begin{equation}
\chi \ge \frac{R-S - w(T-S)} {-(T-R) + w(T-S)}
\label{eq:generous cnd chi g_1}
\end{equation}
and 
\begin{equation}
w > \frac{T-R}{T-S}.
\label{eq:generous cnd w g_1}
\end{equation}
When $w\le (T-R)/(T-S)$, a positive $\chi$ value that satisfies Eq.~\eqref{eq:generous cnd g_1} does not exist. Second, because ${\rm d}g_2/{\rm d}\chi >0$ and $g_2$ is continuous for $\chi\ge 0$,
Eq.~\eqref{eq:generous cnd g_2} is equivalent to
\begin{equation}
\chi \ge \frac{R-S - w(P-S)} {-(T-R) + w(T-P)}
\label{eq:generous cnd chi g_2}
\end{equation}
and 
\begin{equation}
w > \frac{T-R}{T-P}.
\label{eq:generous cnd w g_2}
\end{equation}
When $w\le (T-R)/(T-P)$, a positive $\chi$ value that satisfies Eq.~\eqref{eq:generous cnd g_2} does not exist. Third, because ${\rm d}(g_2/g_1)/{\rm d}\chi > 0$ and $g_2/g_1$ is continuous for $\chi\ge 0$,
Eq.~\eqref{eq:generous cnd g_2/g_1} is equivalent to
\begin{equation}
\chi \ge \frac{T-P - w(T-R)} {-(P-S) + w(R-S)}
\label{eq:generous cnd chi g_2/g_1}
\end{equation}
and 
\begin{equation}
w > \frac{P-S}{R-S}.
\label{eq:generous cnd w g_2/g_1}
\end{equation}
When $w\le (P-S)/(R-S)$, a positive $\chi$ value that satisfies Eq.~\eqref{eq:generous cnd g_2/g_1} does not exist. 

By combining Eqs.~\eqref{eq:generous cnd w g_1}, \eqref{eq:generous cnd w g_2}, and \eqref{eq:generous cnd w g_2/g_1}, we find that a generous strategy exists if and only if $w>w_{\rm c}$, where $w_{\rm c}$ is given by Eq.~\eqref{eq:cnd equalizer w}. Therefore, the threshold $w$ value above which a ZD strategy exists is the same for the equalizer, extortioner, and generous strategy. It should be noted that $w=w_{\rm c}$ is allowed for the equalizer, but not for the extortioner and the generous strategy. When $w>w_{\rm c}$,  
Eq.~\eqref{eq:generous cnd chi g_2} implies Eq.~\eqref{eq:generous cnd chi g_1}, and hence one obtains
\begin{equation}
\chi \ge \chi_{\rm c}(w) \equiv \max \left( \frac{R-S - w(P-S)} {-(T-R) + w(T-P)},
\frac{T-P - w(T-R)} {-(P-S) + w(R-S)} \right).
\label{eq:generous cnd chi final}
\end{equation}
Note that $\chi_{\rm c}(w)>1$ and $\chi_{\rm c}(w)$ decreases as $w (>w_{\rm c})$ increases. Equation~\eqref{eq:generous cnd chi final} implies that
$\lim_{w\to w_{\rm c}+0} \chi_{\rm c}(w) = \infty$ and $\lim_{w\to 1} \chi_{\rm c}(w) = 1$, which are the same asymptotic as the case of the extortioner.





\section{Conclusions}

We analyzed ZD strategies in finitely repeated prisoner's dilemma games with general payoff matrices. Apart from the derivation of  convenient expressions for ZD strategies, the novel results derived in the present article are two-fold. First, we derived the threshold discount factor value, $w_{\rm c}$, above which the ZD strategies exist for three commonly studied classes of ZD strategies, i.e., equalizer, extortioner, and generous strategies. They all share the same threshold value. Similar to the case of the condition for mutual cooperation in direct reciprocity, ZD strategies can exist only when there are sufficiently many rounds.
Second, we showed that the memory-one strategies that impose a linear relationship between the payoff of the two players are either ZD strategies (Eqs.~\eqref{eq:p zd 1} and \eqref{eq:sol-B-1}) or an unconditional strategy (Eq.~\eqref{eq:rrrrr}). The latter class includes the unconditional cooperator and unconditional defector as special cases. Therefore, for finitely repeated prisoner's dilemma games (i.e., $w<1$), we answered affirmatively to the conjecture posed in \cite{Hilbe2013PlosOne-zd}. With a continuity argument, our results also cover the infinite case, by the consideration of the limit $w \to 1$. In other words, if the two payoffs are in a linear relationship for any $w=1-\epsilon$, where $\epsilon\ll 1$, then the payoffs are also on a line as $\epsilon$ goes to $0$. For a similar argument, see Eqs.~(5) and (6) in Ref.~\cite{Hilbe2015GamesEconBehav}. The present results also hold true when the co-player employs a longer-memory strategy, because it is straightforward to apply the proof for the infinite case \cite{Press2012PNAS} to the finite case.

Our analytical approach is different from the previous approaches. Press and Dyson's derivation is based on the linear algebra of matrices \cite{Press2012PNAS}. The proof in Ref.~\cite{Hilbe2013PNAS} considers certain telescoping sums. The approach considered in the present study is more elementary than theirs, i.e., to derive necessary conditions and show that they are sufficient by straightforward calculations.

We mention possible directions of future research. First, we conjecture that the $w_{\rm c}$ value is the same for all ZD strategies because it takes the same value for the three common ZD strategies.
Second, the explicit forms of our solutions (Eqs.~\eqref{eq:final p equalizer} and ~\eqref{eq:p zd 1}) may be useful for exploring features of ZD strategies in finitely repeated games. For example, investigation of evolutionary dynamics and extensions to multiplayer games, which have been examined for infinitely repeated games (see section~\ref{sec:introduction} for references), in the case of finitely repeated games may benefit from the present results.

\appendix

\section{Expression of $\bm u^{\rm eq, 0000}$, $\pi_{Y, 0000}$, $\bm u^{\rm eq, 1111}$, and $\pi_{Y, 1111}$\label{sec:equalizer u and pi}}

By substituting $\bm{q}=(0, 0, 0, 0)$ in Eq.~\eqref{eq:M} and then substituting the obtained $M$ in Eq.~\eqref{eq:u^eq}, we obtain
\begin{align}
\bm u^{\rm eq, 0000} =& \frac{1}{(1-w)(1-wp_{\rm CD}+wp_{\rm DD})} \times\notag\\
& \begin{pmatrix}
(1-w)(1-w p_{\rm CD} + wp_{\rm DD})R + w(1-p_{\rm CC} +wp_{\rm CC}-wp_{\rm CD})P + w(p_{\rm CC}-wp_{\rm CC}+wp_{\rm DD})T\\[6pt]
w(1-p_{\rm CD} )P + (1-w+wp_{\rm DD})T\\[6pt]
(1-w)(1-wp_{\rm CD}+wp_{\rm DD})S+w(1-p_{\rm DC}-wp_{\rm CD}+wp_{\rm DC})P + w(p_{\rm DC}-wp_{\rm DC}+wp_{\rm DD})T\\[6pt]
(1-wp_{\rm CD})P + wp_{\rm DD}T
\end{pmatrix},
\end{align}
which leads to
\begin{equation}
\pi_{Y, 0000} = (1-w)\bm v(0)\bm u^{\rm eq, 0000}.
\end{equation}
Similarly, by substituting $\bm{q}=(1, 1, 1, 1)$ in Eq.~\eqref{eq:M} and then substituting the obtained $M$ in Eq.~\eqref{eq:u^eq}, we obtain
\begin{align}
\bm u^{\rm eq, 1111} =& \frac{1}{(1-w)(1-wp_{\rm CC}+wp_{\rm DC})} \times \notag\\
& \begin{pmatrix}
w(1-p_{\rm CC})S+(1-w+wp_{\rm DC})R\\[6pt]
(1-w)(1-wp_{\rm CC}+wp_{\rm DC})T+w(1-p_{\rm CD} - wp_{\rm CC} + wp_{\rm CD})S + w(p_{\rm CD} - wp_{\rm CD} + wp_{\rm DC})R\\[6pt]
(1-wp_{\rm CC})S+wp_{\rm DC}R\\[6pt]
(1-w)(1-wp_{\rm CC}+wp_{\rm DC})P + w(1-p_{\rm DD}-wp_{\rm CC} + wp_{\rm DD})S + w(p_{\rm DD} + wp_{\rm DC} - wp_{\rm DD})R
\end{pmatrix},
\end{align}
which leads to
\begin{equation}
\pi_{Y, 1111} = (1-w)\bm v(0)\bm u^{\rm eq, 1111}.
\end{equation}

\section{Expression of $\bm u^{\rm zd, 0000}$ and $\bm u^{\rm zd, 1111}$, and four necessary conditions in section~\ref{sub:general cases}\label{app:general cases}}

By substituting $\bm{q}=(0, 0, 0, 0)$ in Eq.~\eqref{eq:M} and then substituting the obtained $M$ in Eq.~\eqref{eq:zd_u}, we obtain
\begin{equation}
\bm u^{\rm zd, 0000}= 
\begin{pmatrix}
\cfrac{-w (1 - p_{\rm CC} +wp_{\rm CC} - wp_{\rm CD})(\chi-1) P + w (p_{\rm CC} - wp_{\rm CC} + wp_{\rm DD})(S-\chi T)}{1 - wp_{\rm CD} + wp_{\rm DD}} + (\chi - 1)\kappa - (1-w)(\chi-1)R\\[20pt]
\cfrac{-w (1-p_{\rm CD})(\chi-1) P + (1 - w+ wp_{\rm DD})(S - \chi T)}
{1 - wp_{\rm CD} + wp_{\rm DD}} + (\chi -1)\kappa \\[20pt]
\cfrac{-w (1 - p_{\rm DC} - wp_{\rm CD} + wp_{\rm DC})(\chi-1) P + w (p_{\rm DC} - wp_{\rm DC} + wp_{\rm DD})(S-\chi T)}{1 - wp_{\rm CD} + wp_{\rm DD}} + (\chi - 1)\kappa + (1 - w)(T-\chi S)\\[20pt]
\cfrac{-(1-w p_{\rm CD})(\chi-1)P+ wp_{\rm DD} (S - \chi T)}
{1 -w p_{\rm CD} + w p_{\rm DD}} + (\chi - 1)\kappa
\end{pmatrix}.
\label{eq:u^0000}
\end{equation}
By substituting $\bm{q}=(1, 1, 1, 1)$ in Eq.~\eqref{eq:M} and then substituting the obtained $M$ in Eq.~\eqref{eq:zd_u}, we obtain
\begin{equation}
\bm u^{\rm zd, 1111}=\begin{pmatrix}
\cfrac{w(1 - p_{\rm CC})(T -\chi S) - (1 - w+wp_{\rm DC})(\chi-1)R}{1 - wp_{\rm CC} + wp_{\rm DC}} +(\chi - 1)\kappa \\[20pt]
\cfrac{w (1-p_{\rm CD} - wp_{\rm CC} + wp_{\rm CD})(T-\chi S) - w (p_{\rm CD} - wp_{\rm CD} + wp_{\rm DC})(\chi-1)R}{1 - wp_{\rm CC} + wp_{\rm DC}} + (\chi - 1)\kappa + (1 - w) (S-\chi T)\\[20pt]
\cfrac{(1-w p_{\rm CC})(T-\chi S) - wp_{\rm DC} (\chi-1) R}{1 - wp_{\rm CC} + wp_{\rm DC}} +(\chi - 1)\kappa \\[20pt]
\cfrac{w(1-p_{\rm DD}-wp_{\rm CC}+wp_{\rm DD})(T-\chi S) - w(p_{\rm DD}+wp_{\rm DC}-wp_{\rm DD})(\chi-1)R}{1 - wp_{\rm CC} + wp_{\rm DC}} + (\chi-1)\kappa - (1-w)(\chi-1)P
\end{pmatrix}.
\label{eq:u^1111}
\end{equation}
Note that the denominator on the right-hand side of Eqs.~\eqref{eq:u^0000} and \eqref{eq:u^1111} is positive.

By substituting Eq.~\eqref{eq:u^0000} in Eq.~\eqref{eq:zd3-1}, we obtain
\begin{align}
  (1-w)p_0 \left\{ (1-wp_{\rm CD} + wp_{\rm DD})\left[-(\chi-1)R - T + \chi S\right] + w(p_{\rm CC}-p_{\rm DC})\left[(\chi-1)P + S - \chi T\right] \right\}\notag\\ +  (1 - wp_{\rm CD} + wp_{\rm DD})\left[(\chi - 1)\kappa+(1 - w) (T-\chi S) \right] + w [-(1 - p_{\rm DC} - wp_{\rm CD} + wp_{\rm DC})(\chi-1)P
  \notag\\ +(p_{\rm DC} - wp_{\rm DC} + wp_{\rm DD})(S-\chi T)] =0.
\label{eq:zd cnd 1}
\end{align}
By substituting Eq.~\eqref{eq:u^0000} in Eq.~\eqref{eq:zd3-2}, we obtain
\begin{align}
 (1-w)p_0 \left[(\chi-1)P+ S -\chi T\right] - (1-wp_{\rm CD})(\chi-1) P + wp_{\rm DD} (S-\chi T) \notag\\
 +  (1-wp_{\rm CD} + wp_{\rm DD})(\chi - 1)\kappa =0.
\label{eq:zd cnd 2}
\end{align}
By substituting Eq.~\eqref{eq:u^1111} in Eq.~\eqref{eq:zd3-1}, we obtain
\begin{align}
  (1-w)p_0 \left[-(\chi-1) R-T +\chi S\right]\notag\\
 + (1-w p_{\rm CC})(T-\chi S) - wp_{\rm DC} (\chi-1) R + (1-w p_{\rm CC} +w p_{\rm DC})(\chi-1)\kappa =0.
\label{eq:zd cnd 3}
\end{align}
By substituting Eq.~\eqref{eq:u^1111} in Eq.~\eqref{eq:zd3-2}, we obtain
\begin{align}
  (1-w)p_0 \left\{(1-wp_{\rm CC} + wp_{\rm DC})\left[(\chi-1)P + S-\chi T\right] + w(p_{\rm CD}-p_{\rm DD})\left[-(\chi-1)R -T + \chi S\right]\right\}\notag\\
   +  (1 - wp_{\rm CC} + wp_{\rm DC})(\chi-1)\left[\kappa - (1 - w)P \right]+ w [(1-p_{\rm DD}-wp_{\rm CC}+wp_{\rm DD})(T-\chi S) 
  \notag\\ - (p_{\rm DD}+wp_{\rm DC}-wp_{\rm DD})(\chi-1)R] =0. &
\label{eq:zd cnd 4}
\end{align}

\section{Derivation of Eq.~\eqref{eq:rrrrr}\label{sec:rrrrr calc}}

In this section, we derive Eq.~\eqref{eq:rrrrr} from Eqs.~\eqref{eq:kappa rrrr} and \eqref{eq:chi rrrr}.

We obtain
\begin{equation}
\bm u^{\rm zd, 1000}= 
\begin{pmatrix}
\cfrac{1}{(1 - w p_{\rm CC}) (1 - w p_{\rm CD} + w p_{\rm DD})}\\
\times \left \{ \left \{-(1 - w) (1 - w p_{\rm CD} + w p_{\rm DD}) R - w^2 (1 - p_{\rm CC}) [1 - (1 - w) p_{\rm DC} - w p_{\rm CD}] P \right \} (\chi-1) \right. \\
\left. + w (1 - p_{\rm CC}) \left \{(1 - w) (1 - w p_{\rm CD} + w p_{\rm DD}) (T - \chi S) + w [(1 - w) p_{\rm DC} + w p_{\rm DD}] (S - \chi T) \right \} \right \}+ (\chi - 1)\kappa \\[20pt]
\cfrac{- w (1-p_{\rm CD})(\chi-1) P + (1 - w+ wp_{\rm DD})(S - \chi T)}
{1 - wp_{\rm CD} + wp_{\rm DD}} + (\chi -1)\kappa \\[20pt]
\cfrac{-w (1 - p_{\rm DC} - wp_{\rm CD} + wp_{\rm DC})(\chi-1) P + w (p_{\rm DC} - wp_{\rm DC} + wp_{\rm DD})(S-\chi T)}{1 - wp_{\rm CD} + wp_{\rm DD}} + (\chi - 1)\kappa + (1 - w)(T-\chi S)\\[20pt]
\cfrac{-(1-w p_{\rm CD})(\chi-1)P+ wp_{\rm DD} (S - \chi T)}
{1 -w p_{\rm CD} + w p_{\rm DD}} + (\chi - 1)\kappa
\end{pmatrix}.
\label{eq:u^1000}
\end{equation}
Note that the denominator on the right-hand side of Eq.~\eqref{eq:u^1000} is positive. By substituting Eq.~\eqref{eq:u^1000} in Eq.~\eqref{eq:zd3-2}, we obtain Eq.~\eqref{eq:zd cnd 2}. By substituting Eq.~\eqref{eq:u^1000} in Eq.~\eqref{eq:zd3-1}, we obtain
\begin{align}
p_0 (1 - w) \left \{ \left \{-(1 - w p_{\rm CD} + w p_{\rm DD}) R + w [1 - (1 - w) p_{\rm DC} - w p_{\rm CD}] P\right \} (\chi-1) \right. \notag \\
\left. - (1 - w) (1 - w p_{\rm CD} + w p_{\rm DD}) (T - \chi S) - w [(1 - w) p_{\rm DC} + w p_{\rm DD}] (S - \chi T)\right \} \notag\\
 + (1 -w p_{\rm CC}) \left \{-w [1 - (1 - w) p_{\rm DC} - w p_{\rm CD}] (\chi-1) P + w [(1 - w) p_{\rm DC} + w p_{\rm DD}] (S - \chi T)\right \} \notag\\
 + (1 - w p_{\rm CC}) (1 - w p_{\rm CD} + w p_{\rm DD}) (\chi - 1) \kappa + (1 - w p_{\rm CC}) (1 - w p_{\rm CD} + w p_{\rm DD}) (1 - w) (T - \chi S) =0.  
\label{eq:zd cnd 5}
\end{align}
Substitution of Eqs.~\eqref{eq:p_DD in terms of p_CD} and \eqref{eq:p_CC in terms of p_DC} in Eq.~\eqref{eq:zd cnd 5} yields either the third entry of Eq.~\eqref{eq:p zd 1} or 
\begin{align}
\frac{(p_0 -p_{\rm DC})(\kappa -R)(1-w)w(\chi-1)\left[(\chi-1)P+S-\chi T\right]}
{\left[T - \kappa + \chi (\kappa - S)\right] \left[\kappa - S + \chi (T - \kappa)\right]}
=0.
\label{eq:cnd rrrr before chi and kappa 1}
\end{align}
The case in which the denominator on the right-hand side of Eq.~\eqref{eq:cnd rrrr before chi and kappa 1} is equal to 0 is covered in Appendices~\ref{sec:denom=0 case 1} and \ref{sec:denom=0 case 2}. We note that $\chi\neq 1$ because $\chi=1$ substituted in Eq.~\eqref{eq:chi rrrr} yields $T=S$, which contradicts Eq.~\eqref{eq:T>R>P>S}. By combining this observation with $0<w<1$, we obtain
\begin{equation}
(p_0 -p_{\rm DC})(\kappa -R)\left[(\chi-1)P+S-\chi T\right] = 0.
\label{eq:cnd rrrr before chi and kappa 2}
\end{equation}
By substituting Eqs.~\eqref{eq:kappa rrrr} and \eqref{eq:chi rrrr} in
Eq.~\eqref{eq:cnd rrrr before chi and kappa 2}, we obtain the following four possible cases: $p_0 = p_{\rm DC}$, $p_0=1$, $p_0 = (R-P)/(T+S-R-P)$, and $p_0 = (T+S-2P)/(T+S-R-P)$.

First, assume that $p_0 = p_{\rm DC}$. By substituting $p_0=p_{\rm DC}$ and Eq.~\eqref{eq:p_0 cnd 1} in Eq.~\eqref{eq:zd cnd 3}, we obtain 
$(p_{\rm CC} - p_{\rm DC})\left[T-\kappa+\chi(\kappa-S)\right] = 0$.
Because we have excluded the case $T-\kappa+\chi(\kappa-S) = 0$, which we deal with in Appendix~\ref{sec:denom=0 case 1}, we obtain $p_{\rm CC}=p_{\rm DC}$. Therefore, we obtain
\begin{equation}
p_0 = p_{\rm CC} = p_{\rm DC}.
\label{eq:p0 for rrrrr case 1-1}
\end{equation}

Second, assume that $p_0=1$. Substitution of $p_0=1$ in Eq.~\eqref{eq:kappa rrrr} yields $\kappa=R$.
Substitution of $p_0=1$ and $\kappa=R$ in Eq.~\eqref{eq:p_CC in terms of p_DC} yields $p_{\rm CC} = 1$.
Substitution of $p_0=1$ in Eq.~\eqref{eq:chi rrrr} yields $\chi= -(R-S)/(T-R)$.
Substitution of $p_0=1$, $\chi= -(R-S)/(T-R)$, and $\kappa=R$ in Eq.~\eqref{eq:zd cnd 2} yields
$(1-p_{\rm CD})(T-S)(R-P)=0$, which implies $p_{\rm CD}=1$. Therefore, $p_0=1$ combined with Eqs.~\eqref{eq:kappa rrrr} and \eqref{eq:chi rrrr} results in 
\begin{equation}
p_0 = p_{\rm CC} = p_{\rm CD} = 1.
\label{eq:p0 for rrrrr case 1-2}
\end{equation}

Third, we note that
\begin{equation}
p_0 \neq \frac{R-P}{T+S-R-P}
\label{eq:p0 neq (R-P)/(T+S-R-P)}
\end{equation}
because combination of $p_0 = (R-P)/(T+S-R-P)$ and $0\le p_0\le 1$ leads to $T+S-R-P>0$ and $2R\le T+S$, and the latter inequality contradicts Eq.~\eqref{eq:2R>T+S}.

Fourth, assume that $p_0 =(T+S-2P)/(T+S-R-P)$. By substituting $p_0 =(T+S-2P)/(T+S-R-P)$ in Eqs.~\eqref{eq:kappa rrrr} and \eqref{eq:chi rrrr}, we obtain $\chi=-(P-S)/(T-P)$ and $\kappa=P$, respectively. Then, we obtain $\kappa-S + \chi(T-\kappa)=0$, which we have decided to deal with later.

To summarize, Eq.~\eqref{eq:cnd rrrr before chi and kappa 2} leads to either
Eq.~\eqref{eq:p0 for rrrrr case 1-1} or \eqref{eq:p0 for rrrrr case 1-2}.

We obtain
\begin{align}
\bm u^{\rm zd, 0001}=& 
(\chi-1)\kappa \begin{pmatrix}
1\\1\\1\\1
\end{pmatrix}
+ \cfrac{1}{(1 + w)(1 - w p_{\rm CD}) + w^2 \left[p_{\rm DC}(1 - p_{\rm DD}) + p_{\rm CC}p_{\rm DD})\right]}
\times\notag\\
& \begin{pmatrix}
\left \{ \left [-1 + w p_{\rm CD} + w^2 (1 - p_{\rm DC}) (1 - p_{\rm DD}) - w^3 (p_{\rm CD}-p_{\rm DC}) (1 - p_{\rm DD}) \right ] R \right. \\
\left.+ w [-1 + (1 - w) p_{\rm CC} + w p_{\rm CD}] P\right \} (\chi - 1) + w \left \{w (1 - p_{\rm DD}) [(1 - w p_{\rm CD}) - (1 - w) p_{\rm CC}] (T - \chi S) \right. \\
\left. + [p_{\rm CC}-w^2 (1 - p_{\rm DD}) (p_{\rm CC} - p_{\rm DC})] (S - \chi T)\right \} \\[20pt]
w (1 - p_{\rm CD}) [-(P + w p_{\rm DD} R) (\chi - 1) + w (1 - p_{\rm DD}) (T - \chi S)] \\+ [1 - w^2 (1 - p_{\rm DC} - p_{\rm CC}p_{\rm DD} + p_{\rm DC}p_{\rm DD})] (S - \chi T)\\[20pt]
w \left[-1 + w p_{\rm CD} + (1 - w)p_{\rm DC} \right] (P+w p_{\rm DD}R) (\chi - 1) \\
+ \left \{1 - w^2 p_{\rm DD} [1 - (1 - w) p_{\rm CC}] - w p_{\rm CD} (1 - w^2 p_{\rm DD})\right \} (T - \chi S) + w [p_{\rm DC} + w^2 (p_{\rm CC} - p_{\rm DC}) p_{\rm DD}] (S - \chi T)\\[20pt]
-\left[(1 - w p_{\rm CD}) P + w p_{\rm DD} (1 - w p_{\rm CD}) R\right] (\chi - 1) + w \left \{(1 - p_{\rm DD}) (1 - w p_{\rm CD}) (T - \chi S) \right. \\
\left. + w (p_{\rm DC} + p_{\rm CC}p_{\rm DD} - p_{\rm DC}p_{\rm DD}) (S - \chi T)\right \}
\end{pmatrix}.
\label{eq:u^0001}
\end{align}
Note that the denominator on the right-hand side of Eq.~\eqref{eq:u^0001} is positive. By substituting Eq.~\eqref{eq:u^0001} in Eq.~\eqref{eq:zd3-2}, we obtain
\begin{align}
(1 - w) p_0 \left \{(P + w p_{\rm DD} R) (\chi - 1) - w (1 - p_{\rm DD}) (T - \chi S) + (1 + w) (S - \chi T)\right \} - [(1 - w p_{\rm CD}) P \notag \\
+  w p_{\rm DD} (1 - w p_{\rm CD}) R] (\chi-1) + w \left \{(1 - p_{\rm DD}) (1 - w p_{\rm CD}) (T - \chi S) + w [p_{\rm DC} + p_{\rm DD} (p_{\rm CC} - p_{\rm DC})] (S - \chi T)\right \} \notag \\
+ \left \{1 + w (1 - p_{\rm CD}) - w^2 [p_{\rm CD} - p_{\rm DC} - p_{\rm DD} (p_{\rm CC} - p_{\rm DC})]\right \} (\chi - 1) \kappa =0.
\label{eq:zd cnd 7}
\end{align}
Substitution of Eqs.~\eqref{eq:p_DD in terms of p_CD} and \eqref{eq:p_CC in terms of p_DC} in Eq.~\eqref{eq:zd cnd 7} yields either the third entry of Eq.~\eqref{eq:p zd 1} or
\begin{align}
\cfrac{1}{[T - \kappa + \chi (\kappa - S)] [\kappa - S + \chi (T - \kappa)]}
\times \left \{ w (\chi - 1)^2 \kappa^2 - [1 - (1 - w)p_0 - w p_{\rm CD}] (\chi - 1) [T - R + \chi (R - S)] P \right. \notag \\
\left. + \left \{w (T - \chi S) + (1 - w) p_0 [T - R + \chi (R - S)]\right \} (S - \chi T) -\left \{- (1 - w p_{\rm CD}) (\chi - 1) R \right. \right. \notag \\
\left. \left. - [1 + w (1 - p_{\rm CD} - \chi)] T + [\chi - w (1 - \chi + p_{\rm CD} \chi)] S \right \} (\chi - 1) \kappa \right \}=0.
\label{eq:cnd rrrr before chi and kappa 3}
\end{align}
We examine the case in which the denominator on the right-hand side of Eq.~\eqref{eq:cnd rrrr before chi and kappa 3} is zero in Appendices~\ref{sec:denom=0 case 1} and \ref{sec:denom=0 case 2}. Therefore, we ignore the denominator and substitute Eqs.~\eqref{eq:kappa rrrr} and \eqref{eq:chi rrrr} in
Eq.~\eqref{eq:cnd rrrr before chi and kappa 3} to obtain
$p_0=p_{\rm CD}$, $p_0 = 0$, $p_0 = (R-P)/(T+S-R-P)$, or $p_0 = (T+S-2P)/(T+S-R-P)$. Among these four possible options, we have excluded $p_0 = (R-P)/(T+S-R-P)$ and $p_0 = (T+S-2P)/(T+S-R-P)$ in the course of the analysis of $\bm u^{\rm zd, 1000}$. 

First, assume that $p_0 = p_{\rm CD}$. By substituting $p_0=p_{\rm CD}$ and Eq.~\eqref{eq:p_0 cnd 2} in Eq.~\eqref{eq:zd cnd 2}, we obtain 
$(p_{\rm CD} - p_{\rm DD})\left[\kappa-S+\chi(T-\kappa)\right] = 0$.
Because we have excluded the case $\kappa-S+\chi(T-\kappa) = 0$, which we deal with in Appendix~\ref{sec:denom=0 case 1}, we obtain $p_{\rm DD}=p_{\rm CD}$. Therefore,
we obtain
\begin{equation}
p_0 = p_{\rm CD} = p_{\rm DD}.
\label{eq:p0 for rrrrr case 2-1}
\end{equation}

Second, assume that $p_0=0$. Substitution of $p_0=0$ in Eq.~\eqref{eq:kappa rrrr} yields $\kappa=P$. Substitution of $p_0=0$ and $\kappa=P$ in Eq.~\eqref{eq:p_DD in terms of p_CD} yields $p_{\rm DD} = 0$.
Substitution of $p_0=0$ in Eq.~\eqref{eq:chi rrrr} yields $\chi= -(T-P)/(P-S)$.
Substitution of $p_0=0$, $\chi= -(T-P)/(P-S)$, and $\kappa=P$ in Eq.~\eqref{eq:zd cnd 3} yields
$w p_{\rm DC} (R-P)=0$, which implies $p_{\rm DC}=0$. Therefore, $p_0=0$ combined with Eqs.~\eqref{eq:kappa rrrr} and \eqref{eq:chi rrrr} results in
\begin{equation}
p_0 = p_{\rm DC} = p_{\rm DD} = 0.
\label{eq:p0 for rrrrr case 2-2}
\end{equation}

A solution must simultaneously satisfy either Eq.~\eqref{eq:p0 for rrrrr case 1-1} or \eqref{eq:p0 for rrrrr case 1-2}, and either Eq.~\eqref{eq:p0 for rrrrr case 2-1} or \eqref{eq:p0 for rrrrr case 2-2}. The combination of Eqs.~\eqref{eq:p0 for rrrrr case 1-1} and \eqref{eq:p0 for rrrrr case 2-1} provides the set of unconditional strategies, i.e., Eq.~\eqref{eq:rrrrr}. The combination of Eqs.~\eqref{eq:p0 for rrrrr case 1-1} and \eqref{eq:p0 for rrrrr case 2-2} provides a subset of the strategies given by Eq.~\eqref{eq:rrrrr}.
The combination of Eqs.~\eqref{eq:p0 for rrrrr case 1-2} and \eqref{eq:p0 for rrrrr case 2-1} also provides a subset of the strategies given by Eq.~\eqref{eq:rrrrr}.
Equations~\eqref{eq:p0 for rrrrr case 1-2} and \eqref{eq:p0 for rrrrr case 2-2} are inconsistent with each other. Therefore, the set of solutions is given by Eq.~\eqref{eq:rrrrr}.

\section{An unconditional strategy is not a ZD strategy unless $R+P=T+S$ \label{sec:rrrrr is not ZD}}

In this section, we show that the unconditional strategy given by Eq.~\eqref{eq:rrrrr} is not a ZD strategy in the sense of \cite{Hilbe2015GamesEconBehav} if $R+P\neq T+S$.

By substituting $p_{\rm DC}=p_{\rm DD}$ in Eq.~\eqref{eq:p zd 2}, we obtain
\begin{equation}
\cfrac{\phi \left[(\chi-1)\kappa+T-\chi S\right]-(1-w)p_0}{w}=\cfrac{\phi(\chi-1)(\kappa-P)-(1-w)p_0}{w},
\label{eq:rrrrr is not ZD 1}
\end{equation}
which leads to
\begin{equation*}
\phi [(\chi-1)\kappa+T-\chi S]=\phi(\chi-1)(\kappa-P).
\end{equation*}
If $\phi=0$, we substitute $\phi=0$ in the expression of $p_{\rm DD}$ in Eq.~\eqref{eq:p zd 2} to obtain $p_{\rm DD}=-(1-w)p_0/w$. This equation holds true if and only if $p_0=p_{\rm DD}=0$. Next, we substitute $\phi=0$ in the expression of $p_{\rm CC}$ in Eq.~\eqref{eq:p zd 2} to obtain $p_{\rm CC}=\left[1-(1-w)p_0\right]/w$. This equation holds true if and only if $p_0=p_{\rm CC}=1$, which contradicts $p_0=0$. Therefore, we obtain $\phi\neq 0$.
Given $\phi\neq 0$, Eq.~\eqref{eq:rrrrr is not ZD 1} implies
\begin{equation}
\chi=-\cfrac{T-P}{P-S}.
\label{eq:rrrrr is not ZD chi 1}
\end{equation}
By setting $p_{\rm CC}=p_{\rm CD}$ in Eq.~\eqref{eq:p zd 2} and using $\phi\neq 0$, we obtain
\begin{equation}
\chi=-\cfrac{R-S}{T-R}.
\label{eq:rrrrr is not ZD chi 2}
\end{equation}
By combining Eqs.~\eqref{eq:rrrrr is not ZD chi 1} and \eqref{eq:rrrrr is not ZD chi 2},
we obtain
\begin{equation}
R+P = T+S.
\label{eq:rrrrr is not ZD R+P=T+S}
\end{equation}

Equation~\eqref{eq:rrrrr is not ZD R+P=T+S} is a sufficient condition for the unconditional strategy to be a ZD because substitution of Eqs.~\eqref{eq:kappa rrrr}, \eqref{eq:chi rrrr},  \eqref{eq:rrrrr is not ZD R+P=T+S} and
\begin{equation}
\phi=-\frac{T-R}{(T-S)(R-P)}
\end{equation}
in Eq.~\eqref{eq:p zd 2} yields Eq.~\eqref{eq:rrrrr}.

\section{Case $\kappa-S + \chi(T-\kappa) = 0$\label{sec:denom=0 case 1}}

In this section, we assume 
\begin{equation}
\kappa-S + \chi(T-\kappa) = 0
\label{eq:denom=0 case 1}
\end{equation}
and derive the set of strategies that satisfy Eq.~\eqref{eq:non-equalizer}.

By substituting Eq.~\eqref{eq:denom=0 case 1} in Eq.~\eqref{eq:zd cnd 2}, we obtain
\begin{equation}
(\chi -1) (\kappa - P) \left[(1-w)p_0 - 1 + w p_{\rm CD}\right] = 0.
\label{eq:denom=0 case 1 translated}
\end{equation}
Equation~\eqref{eq:denom=0 case 1} does not allow $\chi=1$ because substitution of $\chi=1$ in Eq.~\eqref{eq:denom=0 case 1} yields $T=S$, which contradicts Eq.~\eqref{eq:T>R>P>S}. Substitution of $\kappa=P$ in Eq.~\eqref{eq:denom=0 case 1} yields $\chi = - (P-S)/(T-P)$.
Alternatively, if we set $(1-w)p_0 - 1 + w p_{\rm CD} = 0$, we obtain $p_0 = p_{\rm CD}=1$. Therefore, we consider the following two subcases, i.e., subcase (A) specified by
\begin{equation}
\kappa=P
\label{eq:subcase A cnd 1}
\end{equation}
and
\begin{equation}
\chi = - \frac{P-S}{T-P},
\label{eq:subcase A cnd 2}
\end{equation}
and subcase (B) specified by
\begin{equation}
\kappa-S + \chi(T-\kappa) = 0
\label{eq:subcase B cnd 1}
\end{equation}
and
\begin{equation}
p_0 = p_{\rm CD}=1.
\label{eq:subcase B cnd 2}
\end{equation}

\subsection{Subcase (A): $\kappa = P$ and $\chi = - (P-S)/(T-P)$\label{sec:(A)}}

By substituting Eqs.~\eqref{eq:subcase A cnd 1} and \eqref{eq:subcase A cnd 2} in Eq.~\eqref{eq:zd cnd 1}, we obtain
\begin{equation}
\cfrac{(1 - w) [1 - w (p_{\rm CD} - p_{\rm DD})][-p_0 (T + S - R - P) + T + S - 2 P] (T - S)}{T-P}=0.
\end{equation}
Because $T>P>S$, $0<w<1$, and there exists no pair of $p_{\rm CD}$ and $p_{\rm DD}$ ($0\le p_{\rm CD}, p_{\rm DD}\le 1$) that satisfies $p_{\rm CD}-p_{\rm DD}=1/w$, we obtain
\begin{equation}
p_0 (T + S - R - P) =T+S-2P.
\label{eq:A 1}
\end{equation}
If we set $T + S - R - P=0$, we obtain $T+S-2P=R-P>0$, which contradicts Eq.~\eqref{eq:A 1}. Therefore, Eq.~\eqref{eq:A 1} leads to $T + S - R - P \neq 0$, and hence
\begin{equation}
p_0 =\frac{T+S-2P}{T + S - R - P}\,.
\label{eq:A p0}
\end{equation}
If $T+S-R-P>0$, the condition $p_0\le 1$ applied to Eq.~\eqref{eq:A p0} yields $R\le P$, which contradicts Eq.~\eqref{eq:T>R>P>S}. Therefore, we obtain $T+S-R-P<0$ and hence $T+S-2P\le 0$.

By substituting Eqs.~\eqref{eq:subcase A cnd 1} and \eqref{eq:subcase A cnd 2} in Eq.~\eqref{eq:zd cnd 5}, we obtain
\begin{equation}
\cfrac{(1 - w) \left[1 - w (p_{\rm CD} - p_{\rm DD})\right] \left \{p_0 \left[R - (1 - w) (T+S)\right] +  (1 - w p_{\rm CC})(T+S) - \left[ 2 - (1 - 2 w) p_0 - 2 w p_{\rm CC}\right]P\right \}(T - S) }{T-P}=0.
\label{eq:A2}
\end{equation}
Because $1-w(p_{\rm CD}-p_{\rm DD}) > 0$, Eq.~\eqref{eq:A2} implies
\begin{equation}
p_0 \left[R - (1 - w) (T+S)\right] + (1 - w p_{\rm CC})(T+S) - \left[2 - (1 - 2 w) p_0 - 2 w p_{\rm CC}\right]P = 0.
\label{eq:A3}
\end{equation}
Substitution of Eq.~\eqref{eq:A p0} in Eq.~\eqref{eq:A3} yields
\begin{equation}
\cfrac{w\left[-p_{\rm CC} (T + S - R - P) + T + S - 2 P\right](T + S-2P) }{T+S-R-P}=0.
\end{equation}
We will deal with the case $T+S-2P=0$ later in this section. Therefore, by assuming $T+S-2P<0$, we obtain
\begin{equation}
p_{\rm CC}=\cfrac{T+S-2P}{T + S - R - P}.
\label{eq:A-1000}
\end{equation}

By substituting Eqs.~\eqref{eq:subcase A cnd 1} and \eqref{eq:subcase A cnd 2} in Eq.~\eqref{eq:zd cnd 3}, we obtain
\begin{equation}
\cfrac{\left \{(1-w) p_0 (R - S - T)  + w p_{\rm DC} R - w p_{\rm CC} (T + S) - \left[ 2 - (1 - w) p_0 - 2 w p_{\rm CC} + w p_{\rm DC}\right] P+T+S\right \}(T - S)}{T-P}=0.
\label{eq:A4}
\end{equation}
By substituting $p_0=p_{\rm CC}=(T+S-2P)/(T + S - R - P)$ in Eq.~\eqref{eq:A4}, we obtain
\begin{equation}
\cfrac{w\left[-p_{\rm DC} (T + S - R - P) + T + S - 2 P\right](P - R) (T - S) }{(T - P) (T + S - R - P)}=0,
\end{equation}
which leads to
\begin{equation}
p_{\rm DC}=\cfrac{T+S-2P}{T + S - R - P}.
\label{eq:A-eq47}
\end{equation}

By substituting Eqs.~\eqref{eq:subcase A cnd 1} and \eqref{eq:subcase A cnd 2} in Eq.~\eqref{eq:zd cnd 7}, we obtain
\begin{equation}
\cfrac{w\left[1-w p_{\rm CD}-(1-w)p_0\right]\left[-p_{\rm DD}(T+S-R-P)+T+S-2P\right](T - S) }{T-P}=0.
\label{eq:A-0001-cnd}
\end{equation}
If $1-w p_{\rm CD} - (1-w)p_0 = 0$, we obtain $p_0 = p_{\rm CD}=1$, which contradicts Eq.~\eqref{eq:A p0}. Therefore, Eq.~\eqref{eq:A-0001-cnd} implies
\begin{equation}
p_{\rm DD}=\cfrac{T+S-2P}{T+S-R-P}.
\end{equation}

To derive another condition, we use the vector $\bm u$ when player $Y$ adopts the tit-for-tat strategy, i.e., $\bm q = (1, 0, 1, 0)$. This vector, denoted by $\bm u^{\rm zd, 1010}$, is given by
\begin{align}
\bm u^{\rm zd, 1010}=& 
(\chi-1)\kappa \begin{pmatrix}
1\\1\\1\\1
\end{pmatrix}
+ \cfrac{1}{(1-w p_{\rm CC})(1 - w^2 p_{\rm DC}) + w^2 p_{\rm CD} p_{\rm DC} (1 - w) + w (1 + w) p_{\rm DD} (1-w p_{\rm CC})+ w^3 p_{\rm CD} p_{\rm DD}}
\times\notag\\
& \begin{pmatrix}
\left \{ -[1 - w (1 - p_{\rm DD}) - w^2 p_{\rm DC} (1 - p_{\rm CD}) + w^3 (1 - p_{\rm CD}) (p_{\rm DC} - p_{\rm DD})] R \right. \\ 
\left. - w^2 (1 - p_{\rm CC}) (1 - p_{\rm DC}) P\right. \} (\chi-1) + w (1 - p_{\rm CC}) [1 - w (1 - p_{\rm DD})] (T - \chi S)\\
+ w^2 (1 - p_{\rm CC}) [p_{\rm DC} - w (p_{\rm DC} - p_{\rm DD})] (S - \chi T)\\[20pt]
\left \{ -w^2 (1 - p_{\rm DC}) [1 - (1 - w) p_{\rm CD} - w p_{\rm CC}] P - w p_{\rm CD} [1 - w (1 - p_{\rm DD})] R \right \}  (\chi-1)\\
+ w [1 - (1 - w) p_{\rm CD} - w p_{\rm CC}] [1 - w (1 - p_{\rm DD})] (T - \chi S) + (1 - w p_{\rm CC}) [1 - w (1 - p_{\rm DD})] (S - \chi T)\\[20pt]
w \left \{-(1 - p_{\rm DC}) (1 - w p_{\rm CC}) P - w p_{\rm CD} [(1 - w) p_{\rm DC} + w p_{\rm DD}] R\right \} (\chi-1)\\
+ (1 - w p_{\rm CC}) [1 - w (1 - p_{\rm DD})] (T - \chi S) + w [(1 - w) p_{\rm DC} + w p_{\rm DD}] (1 - w p_{\rm CC}) (S - \chi T)\\[20pt]
\left \{ -\{1 - w^2 p_{\rm DC} [1 - p_{\rm CD} (1 - w)] - w p_{\rm CC} (1 - w^2 p_{\rm DC})\} P - w^2 p_{\rm CD} p_{\rm DD} R\right \} (\chi-1)\\
+ w^2 p_{\rm DD} [1 - p_{\rm CD} (1 - w) - w p_{\rm CC}] (T - \chi S) + w p_{\rm DD} (1 - w p_{\rm CC}) (S - \chi T)
\end{pmatrix}.
\label{eq:u^1010}
\end{align}
Note that the denominator on the right-hand side of Eq.~\eqref{eq:u^1010} is positive. By substituting  Eq.~\eqref{eq:u^1010} in Eq.~\eqref{eq:zd3-2}, we obtain
\begin{align}
(1 - w) p_0 \left \{ \left \{-w p_{\rm CD} R + [1 + w (1 - p_{\rm CC}) - w^2 (p_{\rm CC} - p_{\rm CD})] P\right \} (\chi-1) +w [1 - (1 - w) p_{\rm CD}  - w p_{\rm CC}] (T - \chi S)\right. \notag \\
\left. + (1 - w p_{\rm CC}) (S -\chi T)\right \} + \left \{ \{-1 + w^2 p_{\rm DC} [1 - (1 - w) p_{\rm CD}] + w p_{\rm CC} (1 - w^2 p_{\rm DC})\} P - w^2 p_{\rm CD} p_{\rm DD} R\right \} (\chi-1)\notag \\
+ w^2 p_{\rm DD} [1 - (1 - w) p_{\rm CD} - w p_{\rm CC}] (T - \chi S) +  w p_{\rm DD} (1 - w p_{\rm CC}) (S - \chi T)\notag \\
+ \{1 - w^2 p_{\rm DC} + (1 - w) w^2 p_{\rm CD} p_{\rm DC} + w p_{\rm DD} (1 + w + w^2 p_{\rm CD}) - w p_{\rm CC} [1 - w^2 p_{\rm DC} + (1 + w) w p_{\rm DD}]\} (\chi -1) \kappa =0.
\label{eq:cnd 2 zd 0000}
\end{align}
By substituting $\kappa = P$ and $\chi = - (P-S)/(T-P)$ in Eq.~\eqref{eq:cnd 2 zd 0000}, we obtain
\begin{equation}
\cfrac{w [(1-w)p_0+w p_{\rm DD}]\left \{p_{\rm CD}[R-(1-w)(T+S)]+(1-wp_{\rm CC})(T+S)-[2-p_{\rm CD}-2w(p_{\rm CC}-p_{\rm CD})]P\right \}(T - S) }{T-P}=0.
\label{eq:case A cnd leading to pCD}
\end{equation}
By substituting $p_0=p_{\rm CC}=p_{\rm DD}= (T+S-2P)/(T + S - R - P)$ in Eq.~\eqref{eq:case A cnd leading to pCD}, we obtain
\begin{equation}
\cfrac{\left[-p_{\rm CD} (T + S - R - P) + T + S - 2 P\right] \left[-w(T+S-2P)+T+S-R-P\right](T+S-2P) }{(T + S - R - P)^2}=0.
\label{eq:A5}
\end{equation}
If $-w(T+S-2P)+T+S-R-P = 0$, Eq.~\eqref{eq:A p0} implies that $w=1/p_0$, i.e., $w=p_0=1$, which contradicts $0<w<1$. Because we decided to treat the case $T+S-2P=0$ later, Eq.~\eqref{eq:A5}, implies
\begin{equation}
p_{\rm CD}=\frac{T+S-2P}{T+S-R-P}.
\label{eq:A p_CD}
\end{equation}

In sum, we obtain $p_0=p_{\rm CC}=p_{\rm CD}=p_{\rm DC}=p_{\rm DD}= (T+S-2P)/(T + S - R - P)$ if $T+S-2P<0$. Substitution of $p_0$ in Eqs.~\eqref{eq:kappa rrrr} and \eqref{eq:chi rrrr} yields $\chi=-(P-S)/(T-P)$ and $\kappa=P$, respectively, coinciding with the condition for subcase (A). Therefore, 
the strategy $p_0=p_{\rm CC}=p_{\rm CD}=p_{\rm DC}=p_{\rm DD}= (T+S-2P)/(T + S - R - P)$, where $T+S-2P<0$, is a special case of Eq.~\eqref{eq:rrrrr}.

Finally, let us consider the case $T+S-2P=0$. By combining this condition with
Eq.~\eqref{eq:A p0}, we obtain $p_0=0$. By substituting
$T+S-2P=0$ and $p_0=0$ in Eq.~\eqref{eq:A4}, we obtain
$w(R-P)p_{\rm DC}=0$, which implies that $p_{\rm DC}=0$. By substituting
$T+S-2P=0$ and $p_0=0$ in Eq.~\eqref{eq:A-0001-cnd}, we obtain
$(1-w p_{\rm CD})(R-P)p_{\rm DD}=0$, which implies that
$p_{\rm DD}=0$. Because $p_0 = p_{\rm DC} = p_{\rm DD} = 0$, 
the focal player $X$ never uses $p_{\rm CC}$ and $p_{\rm CD}$.
Therefore, $p_0=p_{\rm DC}=p_{\rm DD}=0$ specifies a strategy.
By substituting $p_0=0$ in Eqs.~\eqref{eq:kappa rrrr} and \eqref{eq:chi rrrr} and using $T+S-2P=0$,
we obtain $\chi=-(T-P)/(P-S)=-(P-S)/(T-P)=-1$ and $\kappa=P$, respectively, coinciding with the condition for subcase (A). Therefore, the strategy $p_0 = p_{\rm DC} = p_{\rm DD} = 0$ is a special case of Eq.~\eqref{eq:rrrrr}.

\subsection{Subcase (B): $\kappa - S + \chi (T - \kappa) = 0$ and $p_0=p_{\rm CD}=1$\label{sec:(B)}}

By substituting Eqs.~\eqref{eq:subcase B cnd 1} and \eqref{eq:subcase B cnd 2}
in Eq.~\eqref{eq:zd cnd 1}, we obtain
\begin{equation}
(1 - w) (\chi - 1) \left[w p_{\rm DD} (\kappa - R) - w p_{\rm CC} (\kappa - P) + w (R - P) + \kappa - R\right]=0.
\label{eq:B1}
\end{equation}
Note that $0<w<1$. Because $\chi=1$ is inconsistent with $\kappa - S + \chi (T - \kappa) = 0$, Eq.~\eqref{eq:B1} yields
\begin{equation}
p_{\rm DD}=\cfrac{w p_{\rm CC} (\kappa - P) - w (R - P) - (\kappa - R)}{w(\kappa-R)}
\end{equation}
provided that $\kappa\neq R$. We will deal with the case $\kappa=R$ later in this section.
By substituting Eqs.~\eqref{eq:subcase B cnd 1} and \eqref{eq:subcase B cnd 2}
in Eq.~\eqref{eq:zd cnd 3}, we obtain
\begin{equation}
(\chi - 1) \left \{w p_{\rm DC} (\kappa - R) - w p_{\rm CC} (\chi + 1) (\kappa - T) + w \left[ R - (\chi + 1) T + \chi \kappa\right] + \kappa - R\right \}=0,
\label{eq:B2}
\end{equation}
which yields
\begin{equation}
p_{\rm DC}=\cfrac{w p_{\rm CC} (\chi + 1) (\kappa - T) - w \left[R - (\chi + 1) T + \chi \kappa\right] - (\kappa - R)}{w(\kappa-R)}
\end{equation}
provided that $\kappa\neq R$. Therefore, we obtain
\begin{equation}
\bm p = \left(p_{\rm CC}, \; 1, \; \cfrac{w p_{\rm CC} (\chi + 1) (\kappa - T) - w [R - (\chi + 1) T + \chi \kappa] - (\kappa - R)}{w(\kappa-R)}, \; \cfrac{w p_{\rm CC} (\kappa - P) - w (R - P) - (\kappa - R)}{w(\kappa-R)} \right), p_0=1,
\label{eq:sol-B-1 dupl} 
\end{equation}
i.e., Eq.~\eqref{eq:sol-B-1}, as a necessary condition for the linear relationship between the payoff of the two players, i.e., Eq.~\eqref{eq:non-equalizer}.

To verify that Eq.~\eqref{eq:sol-B-1} is sufficient, we substitute Eq.~\eqref{eq:sol-B-1} (i.e., Eq.~\eqref{eq:sol-B-1 dupl}) in Eq.~\eqref{eq:zd_u} to obtain
\begin{equation}
\bm u^{\rm zd} = \left(0, \; 0, \; -\cfrac{(1 - w) (\chi - 1) (\kappa - R)}{w (1-p_{\rm CC})}, \; -\cfrac{(1 - w) (\chi - 1) (\kappa - R)}{w (1-p_{\rm CC})} \right),
\label{eq:case B u^zd}
\end{equation}
which is independent of $\bm q$. By combining Eqs.~\eqref{eq:x(0)}, \eqref{eq:case B u^zd}, and $p_0=1$, we obtain $\bm v(0) \bm u^{\rm zd} = 0$, i.e., Eq.~\eqref{eq:zd_xu}.
Therefore, Eq.~\eqref{eq:sol-B-1} is a solution that satisfies Eq.~\eqref{eq:non-equalizer}.
 
The strategy given by Eq.~\eqref{eq:sol-B-1} is expressed in the form of Eq.~\eqref{eq:p zd 2} if we set
$\phi=- w (1-p_{\rm CC})/\left[(\kappa-R)(\chi-1)\right]$ (and use $\kappa -S+\chi(T-\kappa)=0$ and $p_0 = 1$). As an example, we consider the repeated PD game defined by  $R=3$, $T=5$, $S=-2$, $P=1$, and $w=0.8$. We set $\kappa=2$. Because this solution requires $\kappa-S+\chi(T-\kappa)=0$ (Eq.~\eqref{eq:subcase B cnd 1}), we obtain $\chi=-4/3$. If we set $p_{\rm CC}=0$, we obtain $\bm p = \left(0, \; 1, \; 3/4, \; 3/4 \right)$ and $p_0=1$. This solution cannot be represented in the form of Eq.~\eqref{eq:p zd 1} because Eq.~\eqref{eq:p zd 1} requires $\kappa-S+\chi(T-\kappa)\neq 0$. Consistent with this example, Eq.~\eqref{eq:p zd 2} combined with $\phi=- w (1-p_{\rm CC})/\left[(\kappa-R)(\chi-1)\right]$, $\kappa -S+\chi(T-\kappa)=0$, and $p_0 = 1$ yields $\chi<0$. This can be shown as follows. By substituting $\kappa -S+\chi(T-\kappa)=0$ and $p_0 = 1$ in Eq.~\eqref{eq:p zd 2}, we obtain
\begin{equation}
\bm{p}=\begin{pmatrix}
1-\frac{\phi (\chi-1)(R-\kappa)}{w}\\[1em]
1\\[1em]
1 - \frac{1}{w} + \frac{\phi \left[(\chi-1)\kappa+T-\chi S\right]}{w}\\[1em]
1 - \frac{1}{w} + \frac{\phi(\chi-1)(\kappa-P)}{w}
\end{pmatrix}.
\label{eq:p zd 2 with more}
\end{equation}
Because $p_{\rm CC}\le 1$ must hold true in Eq.~\eqref{eq:p zd 2 with more}, we obtain
\begin{equation}
\phi (\chi-1)(R-\kappa) \ge 0.
\label{eq:prove chi<0 1}
\end{equation}
Because $p_{\rm DD}\ge 0$ must hold true in Eq.~\eqref{eq:p zd 2 with more}, we obtain
\begin{equation}
\phi(\chi-1)(\kappa-P) \ge 0.
\label{eq:prove chi<0 2}
\end{equation}
Given $\phi(\chi-1)\neq 0$ (section~\ref{sub:general cases}) and $R>P$, we find that $P\le \kappa\le R$ must hold true for Eqs.~\eqref{eq:prove chi<0 1} and \eqref{eq:prove chi<0 2} to be simultaneously satisfied.
Therefore, using $\kappa -S+\chi(T-\kappa)=0$ we obtain $\chi = - (\kappa-S)/(T-\kappa) < 0$.

Finally, let us consider the case $\kappa=R$. By substituting $\kappa=R$ in Eq.~\eqref{eq:B1}, we obtain $w p_{\rm CC} (R - P)  = w (R - P)$, which implies that $p_{\rm CC}=1$.
By combining this result with Eq.~\eqref{eq:subcase B cnd 2}, we obtain $p_0=p_{\rm CC}=p_{\rm CD}=1$, which implies that player X never uses $p_{\rm DC}$ and $p_{\rm DD}$.
Therefore, $p_0=p_{\rm CC}=p_{\rm CD}=1$ specifies a strategy.
By substituting $p_0=1$ in Eqs.~\eqref{eq:kappa rrrr} and \eqref{eq:chi rrrr}, we obtain $\chi=-(R-S)/(T-R)$ and $\kappa=R$, respectively, and the former equality coincides with 
Eq.~\eqref{eq:subcase B cnd 2} when $\kappa=R$. Therefore, the strategy $p_0 = p_{\rm CC} = p_{\rm CD} = 1$ is a special case of Eq.~\eqref{eq:rrrrr}.

\section{Case $T-\kappa + \chi(\kappa-S) = 0$\label{sec:denom=0 case 2}}

In this section, we assume 
\begin{equation}
T-\kappa + \chi(\kappa-S) = 0
\label{eq:denom=0 case 2}
\end{equation}
and derive the set of strategies that satisfy Eq.~\eqref{eq:non-equalizer}.

By substituting Eq.~\eqref{eq:denom=0 case 2} in Eq.~\eqref{eq:zd cnd 3}, we obtain
\begin{equation}
(\chi -1) (\kappa - R) \left[(1-w)p_0 + w p_{\rm DC}\right] = 0.
\label{eq:denom=0 case 2 translated}
\end{equation}
Equation~\eqref{eq:denom=0 case 2} does not allow $\chi=1$ because substitution of $\chi=1$ in Eq.~\eqref{eq:denom=0 case 2} yields $T=S$, which contradicts Eq.~\eqref{eq:T>R>P>S}. Substitution of
$\kappa=R$ in Eq.~\eqref{eq:denom=0 case 2} yields $\chi = - (T-R)/(R-S)$.
Alternatively, if we set $(1-w)p_0 + w p_{\rm DC} = 0$, we obtain $p_0 = p_{\rm DC}=0$. Therefore, we consider the following two subcases, i.e., subcase (C) specified by
\begin{equation}
\kappa=R
\label{eq:subcase C cnd 1}
\end{equation}
and
\begin{equation}
\chi = - \frac{T-R}{R-S},
\label{eq:subcase C cnd 2}
\end{equation}
and subcase (D) specified by
\begin{equation}
T-\kappa + \chi(\kappa-S) = 0
\label{eq:subcase D cnd 1}
\end{equation}
and
\begin{equation}
p_0 = p_{\rm DC}=0. 
\label{eq:subcase D cnd 2}
\end{equation}

\subsection{Subcase (C): $\kappa = R$ and $\chi = - (T-R)/(R-S)$\label{sec:(C)}}

By substituting Eqs.~\eqref{eq:subcase C cnd 1} and \eqref{eq:subcase C cnd 2} in Eq.~\eqref{eq:zd cnd 4}, we obtain
\begin{equation}
\cfrac{(1 - w) \left[1 - w (p_{\rm CC} -  p_{\rm DC})\right]\left[- p_0 (T+S-R-P)+R-P \right](T-S) }{R-S}=0.
\label{eq:cnd case C}
\end{equation}
Equation~\eqref{eq:cnd case C} does not hold true because
$0 < w < 1$,
$(R-P) - p_0 (T+S-R-P)\neq 0$ due to Eq.~\eqref{eq:p0 neq (R-P)/(T+S-R-P)}, and $1 - w (p_{\rm CC} -  p_{\rm DC}) > 0$. Therefore there is no solution in this case.

\subsection{Subcase (D): $T-\kappa +\chi(\kappa-S)=0$ and $p_0=p_{\rm DC}=0$\label{sec:(D)}}

By substituting Eqs.~\eqref{eq:subcase D cnd 1} and \eqref{eq:subcase D cnd 2} in
Eq.~\eqref{eq:zd cnd 1}, we obtain
\begin{equation}
w (\chi - 1) \left[w p_{\rm DD} (\kappa - S) (\chi + 1) + (1-w p_{\rm CD}) (\kappa - P)\right]=0.
\label{eq:case D cnd 1}
\end{equation}
We obtain $\chi\neq 1$ because $\chi = 1$ substituted in Eq.~\eqref{eq:subcase D cnd 1} yields $T=S$, which contradicts Eq.~\eqref{eq:T>R>P>S}. Therefore, Eq.~\eqref{eq:case D cnd 1} implies
\begin{equation}
p_{\rm CD}=\cfrac{w p_{\rm DD} (\chi + 1) (\kappa - S) + \kappa - P}{w(\kappa-P)}
\label{eq:case D pCD}
\end{equation}
provided that $\kappa\neq P$.
We will deal with the case $\kappa=P$ later in this section.
By substituting Eqs.~\eqref{eq:subcase D cnd 1} and \eqref{eq:subcase D cnd 2} in
Eq.~\eqref{eq:zd cnd 4}, we obtain
\begin{equation}
(\chi - 1) (1 - w) \left[w p_{\rm DD} (\kappa - R) + (1- w p_{\rm CC}) (\kappa - P) \right]=0.
\label{eq:case D cnd 2}
\end{equation}
Because $0<w<1$ and $\chi\neq 1$, we obtain
\begin{equation}
p_{\rm CC}=\cfrac{w p_{\rm DD} (\kappa-R) + \kappa - P}{w(\kappa-P)}
\label{eq:case D pCC}
\end{equation}
provided that $\kappa\neq P$. Therefore, we obtain
\begin{equation}
\bm p = \left(\cfrac{w p_{\rm DD} (\kappa-R) + \kappa - P}{w(\kappa-P)}, \; \cfrac{w p_{\rm DD} (\chi + 1) (\kappa - S) + \kappa - P}{w(\kappa-P)}, \; 0, \; p_{\rm DD} \right), \; p_0=0,
\label{eq:sol p0pDC}
\end{equation}
where $0\le p_{\rm DD}\le 1$ is a necessary condition for the linear relationship between the payoff of the two players, i.e., Eq.~\eqref{eq:non-equalizer}. In fact,
we substitute $p_{\rm CD}$ given by Eq.~\eqref{eq:sol p0pDC} in
$p_{\rm CD}$ given by Eq.~\eqref{eq:p zd 1} and use Eqs.~\eqref{eq:subcase D cnd 1} and \eqref{eq:subcase D cnd 2} to find that $p_{\rm CC}$, $p_{\rm DC}$, $p_{\rm DD}$ given by Eq.~\eqref{eq:p zd 1} coincide with those given by Eq.~\eqref{eq:sol p0pDC}. Therefore, Eq.~\eqref{eq:sol p0pDC} is a special case of ZD strategies given by Eq.~\eqref{eq:p zd 1}.

Finally, let us consider the case $\kappa=P$. By substituting $\kappa=P$ in Eq.~\eqref{eq:case D cnd 2}, we obtain $w p_{\rm DD}(R-P)=0$, which implies that $p_{\rm DD}=0$. By combining this result with Eq.~\eqref{eq:subcase D cnd 2}, we obtain $p_0=p_{\rm DC}=p_{\rm DD}=0$, which implies that player $X$ never uses $p_{\rm CC}$ and $p_{\rm CD}$. Therefore, $p_0=p_{\rm DC}=p_{\rm DD}=0$ specifies a strategy. By substituting $p_0=0$ in Eqs.~\eqref{eq:kappa rrrr} and \eqref{eq:chi rrrr}, we obtain $\chi=-(T-P)/(P-S)$ and $\kappa=P$, respectively, and the former equality coincides with Eq.~\eqref{eq:case D cnd 1} when $\kappa=P$. Therefore,
the strategy $p_0=p_{\rm DC}=p_{\rm DD}=0$ is a special case of Eq.~\eqref{eq:rrrrr}.

\section{Minimum discount rate for $\chi<0$}

\subsection{ZD strategies with $\kappa=P$\label{app:extortioner chi<0}}

Let us consider Eq.~\eqref{eq:extortioner bm p with phi and chi} under $\phi<0$ and $\chi<1$.
In this case, we obtain Eqs.~\eqref{eq:1/phi 1}, \eqref{eq:1/phi 2}, and \eqref{eq:1/phi 3}, but with all the inequalities flipped (i.e., $\ge$ instead of $\le$). Then, we obtain
\begin{align}
\frac{(\chi-1)\frac{R-P}{P-S}} {\frac{1}{w}} \ge& \frac{1+\chi\frac{T-P}{P-S}} {\frac{1}{w}-1},
\label{eq:phi exists 1 chi<0}\\
\frac{1+\chi\frac{T-P}{P-S}} {\frac{1}{w}} \ge& \frac{(\chi-1)\frac{R-P}{P-S}} {\frac{1}{w}-1},
\label{eq:phi exists 2 chi<0}\\
\chi + \frac{T-P}{P-S} \ge& \frac{(\chi-1)\frac{R-P}{P-S}} {\frac{1}{w}-1},
\label{eq:phi exists 3 chi<0}\\
\chi + \frac{T-P}{P-S} \ge& \frac{1+\chi\frac{T-P}{P-S}} {\frac{1}{w}-1}.
\label{eq:phi exists 4 chi<0}
\end{align}
Equations~\eqref{eq:phi exists 1 chi<0}--\eqref{eq:phi exists 4 chi<0} yield
\begin{align}
\chi \le& - \frac{P-S + (1-w)(R-P)} {T-R+w(R-P)} < 0,
\label{eq:chi extortioner cnd 1 chi<0}\\
\left[(R-P)-(1-w)(T-P)\right]\chi \le& R-P+(1-w)(P-S),
\label{eq:chi extortioner cnd 2 chi<0}\\
\left[w(R-P)-(1-w)(P-S)\right]\chi \le& w(R-P)+(1-w)(T-P),
\label{eq:chi extortioner cnd 3 chi<0}\\
\left[-(P-S)+w(T-S)\right]\chi \le& -w(P-S)+(1-w)(T-P),
\label{eq:chi extortioner cnd 4 chi<0}
\end{align}
respectively.
When $w$ is sufficiently large, the coefficients of $\chi$ on the left-hand sides of
Eqs.~\eqref{eq:chi extortioner cnd 2 chi<0}, \eqref{eq:chi extortioner cnd 3 chi<0}, and
\eqref{eq:chi extortioner cnd 4 chi<0} are positive. In this situation, 
Eqs.~\eqref{eq:chi extortioner cnd 1 chi<0}--\eqref{eq:chi extortioner cnd 4 chi<0} are satisfied by a sufficiently negative large $\chi (<0)$. This result is consistent with the previously obtained result \cite{Hilbe2015GamesEconBehav}.

\subsection{ZD strategies with $\kappa=R$\label{app:generous chi<0}}

In this section, we examine Eqs.~\eqref{eq:generous cnd g_1}, \eqref{eq:generous cnd g_2}, and \eqref{eq:generous cnd g_2/g_1} under the assumption that $\chi<0$.
First, because
${\rm d}g_2/{\rm d}\chi >0$, $g_2$ is discontinuous at $\chi = -(R-S)/(T-R)$, and $g_2<0$ for
$-(R-S)/(T-R)<\chi<0$, Eq.~\eqref{eq:generous cnd g_2} is equivalent to
\begin{equation}
\chi < -\frac{R-S}{T-R}
\label{eq:generous cnd chi g_2 chi<0 1}
\end{equation}
if $w\ge (T-R)/(T-P)$ and
\begin{equation}
\frac{R-S - w(P-S)} {-(T-R) + w(T-P)} < \chi < -\frac{R-S}{T-R}
\label{eq:generous cnd chi g_2 chi<0 2}
\end{equation}
if $w< (T-R)/(T-P)$.
Second, using Eq.~\eqref{eq:generous cnd chi g_2 chi<0 1},
${\rm d}g_1/{\rm d}\chi > 0$, and that $g_1$ is discontinuous at $\chi = -(R-S)/(T-R)$,
we find that Eq.~\eqref{eq:generous cnd chi g_2 chi<0 1} implies Eq.~\eqref{eq:generous cnd g_1}
if $w\ge (T-R)/(T-S)$ and that Eq.~\eqref{eq:generous cnd g_1} is equivalent to
\begin{equation}
\frac{R-S - w(T-S)} {-(T-R) + w(T-S)} < \chi < -\frac{R-S}{T-R}
\label{eq:generous cnd chi g_1 chi<0 2}
\end{equation}
if $w < (T-R)/(T-S)$.
Third, because ${\rm d}(g_2/g_1)/{\rm d}\chi > 0$, $g_2/g_1$ is discontinuous at $\chi = -(T-R)/(R-S)$, and $g_2/g_1 < 0$ for $- (T-R)/(R-S) < \chi < 0$, Eq.~\eqref{eq:generous cnd g_2/g_1} is equivalent to
\begin{equation}
\chi < - \frac{T-R}{R-S}
\label{eq:generous cnd chi g_2/g_1 chi<0 1}
\end{equation}
if $w \ge (P-S)/(R-S)$ and
\begin{equation}
\frac{T-P - w(T-R)} {-(P-S) + w(R-S)} \le \chi < - \frac{T-R}{R-S}
\label{eq:generous cnd chi g_2/g_1 chi<0 2}
\end{equation}
if $w< (P-S)/(R-S)$.

To summarize these results, if $w\ge w_{\rm c}$, generous strategies with
\begin{equation}
\chi < \min \left( -\frac{R-S}{T-R}, -\frac{T-R}{R-S}\right) < -1
\label{eq:generous cnd chi chi<0 for large w}
\end{equation}
exist because Eq.~\eqref{eq:generous cnd chi chi<0 for large w} yields Eqs.~\eqref{eq:generous cnd g_1}, \eqref{eq:generous cnd g_2}, and \eqref{eq:generous cnd g_2/g_1}. This result is consistent with the previously obtained results \cite{Hilbe2015GamesEconBehav}. Note that we have used Eq.~\eqref{eq:2R>T+S} to derive the last inequality in Eq.~\eqref{eq:generous cnd chi chi<0 for large w}.
Even if $w<w_{\rm c}$, negative $\chi$ values that satisfy all the conditions, i.e.,
the set of equations out of Eqs.~\eqref{eq:generous cnd chi g_2 chi<0 1},
\eqref{eq:generous cnd chi g_2 chi<0 2},
\eqref{eq:generous cnd chi g_1 chi<0 2},
\eqref{eq:generous cnd chi g_2/g_1 chi<0 1}, and
\eqref{eq:generous cnd chi g_2/g_1 chi<0 2},
corresponding to the given value of $w$, may exist.

\section*{Acknowledgment}
We acknowledge Christian Hilbe, Shun Kurokawa, and Kohei Tamura for valuable comments on the manuscript.
G.I. acknowledges the support by HAYAO NAKAYAMA Foundation for Science \& Technology and Culture.
N.M. acknowledges the support provided through JST, CREST, Japan (No. JPMJCR1304), and JST, ERATO, Kawarabayashi Large Graph Project, Japan (JPMJER1201).


\end{document}